# Experiments in Engineering Physics

## Bachelor of Engineering and Technology

**Prepared by,**

**P. Kulkarni (M.Sc. Ph.D.)**

**prasanna1609@gmail.com**

# PREFACE

Experiments performed in the Physics Laboratory play a significant role in understanding the concepts taught in the theory. A good accompanying laboratory manual serves as a concise guideline which students can use to complete the experiments without having to refer to several reference books on the subject. A thorough study of the manual prior to the experiment helps the student to immediately start with the performance in the laboratory.

The general practice in several universities for the conduct of experimental laboratory class has been to enable students take observations and allow the submission in one week time. However, the observations do not complete the experiment and serve as only one part of learning in the measurement of the physical quantities in the laboratory. The calculations and the submission of the journal before the end of the experimental turn should be an integral part of the laboratory class. With this motivation a scheme is suggested for the conduct of the laboratory class.

Several initiatives are proposed to achieve the above objectives in this carefully prepared manual for the experiments in the Engineering Physics course. The manual focuses on the consistency of the approach towards experimentation and scientific reporting rather than the accuracy in the individual experiment and its technical details. Author hopes that the students will appreciate the simplicity of the manual and find it useful.

The manual comprises views and suggestions of several expert teachers. In particular, the author acknowledges Prof. S. S. Major, IIT Bombay, Mumbai and Prof. S. Jain, PIET Nagpur. Author would also like to acknowledge more than 400 students at St. Francis de Sales College Nagpur, VRCE Nagpur and Physics Department, IIT Bombay, Mumbai. Appendix figure of CRO has been used from the public domain.

**AUTHOR**

**CONTENTS**

## Scheme of experiments

Students are required to perform 4 experiments in the Lab I and 4 experiments in Lab II. Each regular experimental turn of 2 hours consists of Performance (1 ½ hrs) and Submission (30 min). In the first week of the semester, the students will be introduced to the laboratory practices with on-board explanation followed by a detailed demonstration of each experiment. Manuals will be distributed to the students via e-form, e.g., website, email as well as 1 printed copy. Students are required to either download the manuals from the website and print on their own or get it copied. Each student is required to have his own copy of the manual and preserve it for the entire semester.

### Performance of the experiments

Students are required to **perform** and **submit** at the end of each experimental turn on the same day. Students will do the experiment in groups of 2 or 3 each. They should go to the experimental set-up with the journal, calculator and graph immediately at the start of the experimental turn. Students need to come prepared for the experimental turn, both with the understanding of the basic theory and the experiment. Only the written journal pages, Graph papers, Calculator, Pencil and Rubber will be allowed to the students during the performance.

## Experimental Journals

The following are the guidelines to the students for the performance and submissions. Students should write the journals before coming for the experimental turn. Right side of the journal should be written by pen and the left side of the journal should be written only by pencil.

1. Journal pages are to be hand written in the following steps.

**Right Side of the Journal**: Name of Experiment, Aim, Apparatus, Formulae, Procedure, Results (numbers with pencil), Conclusions, Precautions.

**Left Side of the Journal**: Name of Experiment, Aim, Figure, Formulae, Observation Tables, Calculations, Results.

2. On the day of performance, Students are advised to check the circuit diagram or figures and formulae, ensure the connections, the experimental set-up and start taking the observations.

3. Faculty in-charge will take a round and visit each experimental set-up in first 30 minutes and ensure that the experiment has started. Faculty in-charge will keep the help to a minimum and only to the extent of the operation of the instruments for the measurement parameters. Faculty in-charge can just make few suggestions for smooth experimenting and completion of the observation table.

4. Faculty in-charge will take second round to check the observation table of each group and sign one or two observations. Faculty in-charge may help, however, 5 marks may be deducted if the student requires more than enough help due to non-preparation. Few questions will be asked about the experimental part, formulae, calculations etc and marks would be noted for each student.

5. At the end of 1 ½ hrs, students must stop the experiment, and start with the graph, calculations and results. Each student is required to perform this task individually without consultation with the group members or other students in the batch.

6. Faculty in-charge will make the final round with the students to ensure that students are able to manage the calculations. Faculty in-charge will sign the graph paper of each student.

7. At the end of the turn students will submit their experiment (the journal pages) in the laboratory, which they can collect next day or anytime till next week. No journal correction or verification of marks will be done in the experimental turn. Students will be informed about the next experiment to be prepared.

8. In last 10 minutes, students are advised to verify the experimental set-up of the next experiment.

**Marking Distributions:** Total marks for each experiment 40 Marks

| | |
|---|---|
| 1) Journal write-up | 10 Marks |
| 2) Experimental understanding viva | 10 Marks |
| 3) Observations, graphs of the submitted journal | 10 Marks |
| 4) Calculations, results of the submitted journal | 10 Marks |

**Experiment No 1**

## V-I Characteristics of the Photocell

**Aim**:

1. To study the V-I characteristics of the photocell in the Forward Bias mode.

2. To determine the work function of the cathode material of the photocell using the Reverse Bias mode.

**Apparatus**: Photocell, Regulated DC power supply, Sodium vapor lamp, Ballistic galvanometer, Tap key, Digital voltmeter, Digital Milli-voltmeter, Connecting wires.

**Circuit Diagrams:**

**(a) Forward Bias**

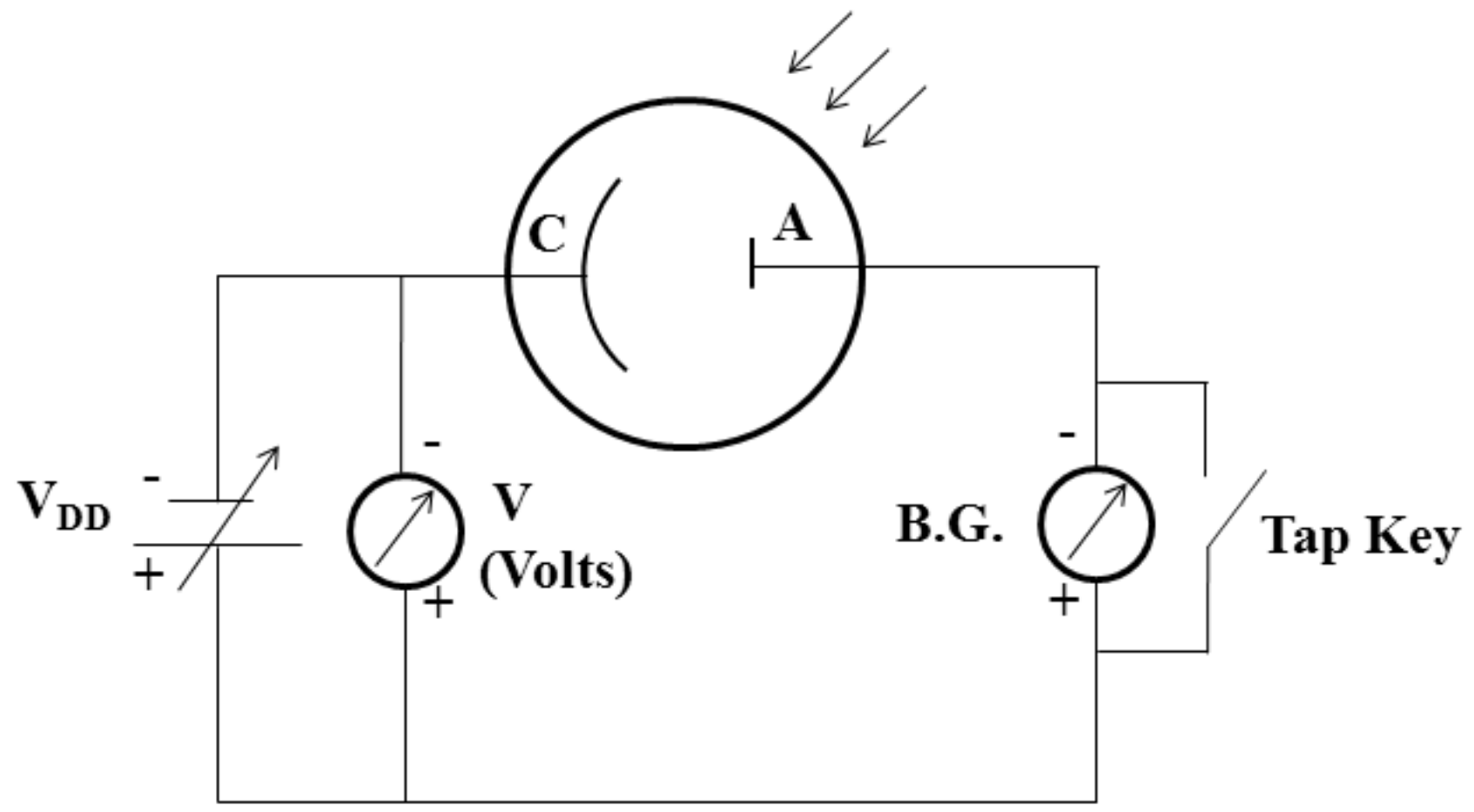

**(b) Reverse Bias**

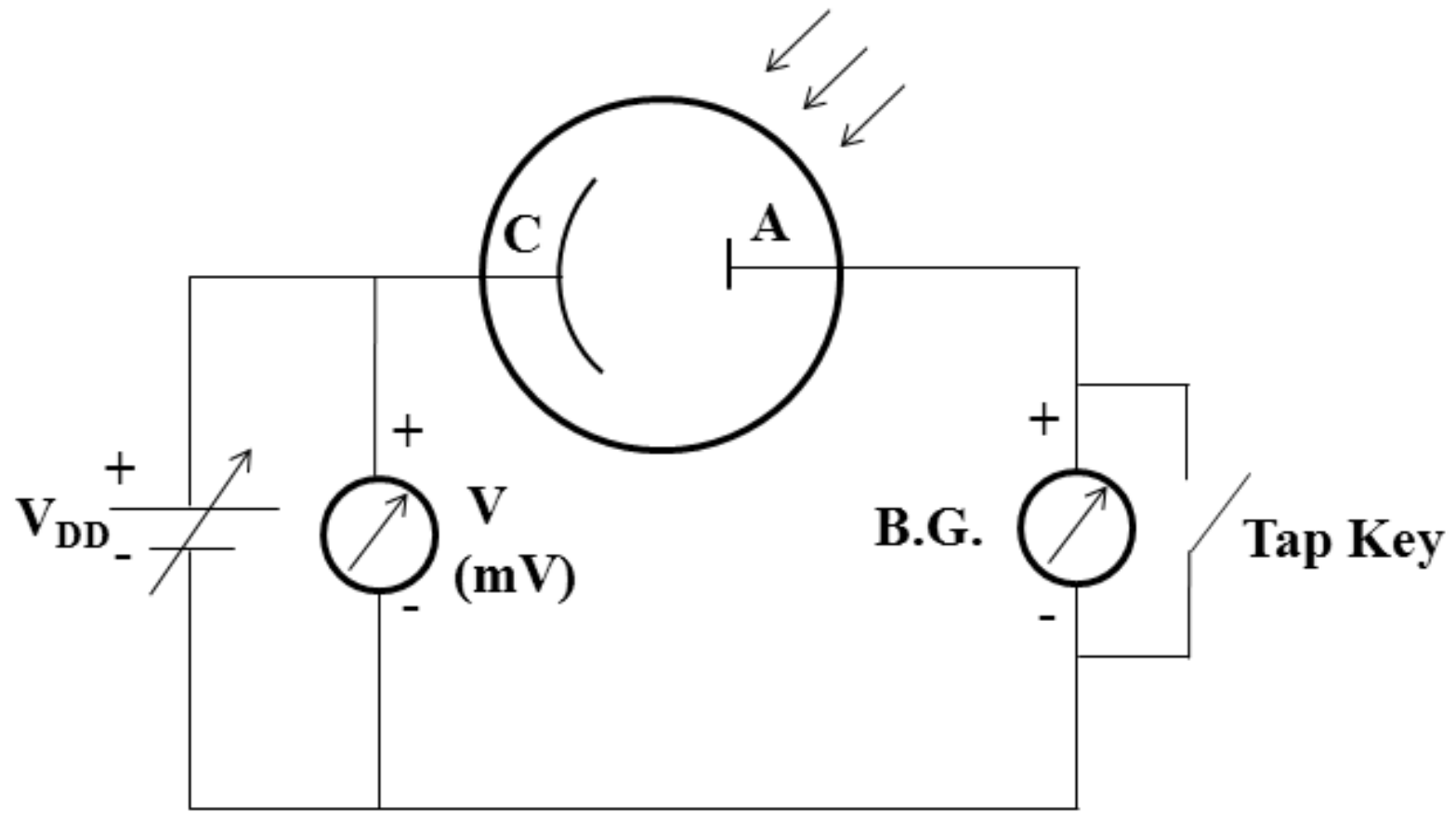

**Formula:** Work function, $\Phi = hc/\lambda - eV_s$ electron-Volts

Where, h = Plank's constant

c = velocity of light

$\lambda$ = wavelength of the sodium light

$V_s$ = stopping potential

**Procedure**:

**1. Forward Bias characteristics of the photocell**

(a) Circuit connections

(i) Connect the regulated voltage source to the given photocell (enclosed in the wooden case and its construction can be seen through the window) with the ballistic galvanometer in series. Connect the digital voltmeter in parallel to the regulated voltage source. Refer to the given circuit diagram (a).

(ii) Ensure the potentiometer on the voltage source in the zero position before switching on the power supply.

(iii) Note the accuracy of the digital voltmeter.

(b) Adjusting the reflection from the mirror of the ballistic galvanometer

(i) Keep the photocell in front of the source window so that the sodium light illuminates and enters the wooden case.

(ii) Move the photocell away from the source to increase the distance between them and to ensure that the intensity of the light reaching the photocell is minimum. This distance would be about 30 cm from the source window.

(iii) Switch ON the regulated power supply and confirm the near zero reading in the digital voltmeter for the minimum position of the potentiometer.

(iv) Note the reflected spot on the graduated scale. Adjust the scale height or lateral position to middle the location of the spot on the scale.

(v) Gradually increase the forward voltage across the photocell by rotating the potentiometer (coarse) on the regulated voltage source clockwise till the digital voltmeter reads 7 volts. Use the coarse knob on the voltage source.

(vi) Keeping the voltage constant, move the photocell closer to the source window while monitoring the movement of the spot on the scale. Adjust the distance till the spot rests at about 20 cm on the scale. Keep the photocell fixed at this position throughout the experiment.

(c) V-I characteristics of the photocell

(i) Note the calibration of the current with respect to the deflection of the galvanometer.

(ii) Decrease the forward voltage of the photocell by rotating the potentiometer (coarse) anticlockwise from 7 volts down to the zero position.

(iii) Note 12 readings of the voltage ($V_{FB}$) and the corresponding deflection (φ) of the spot on the scale. Follow the table 1 to choose the voltage intervals. Calculate the current ($I_{FB}$) through the photocell using the calibration.

(iv) Ensure the minimum position of the potentiometer at the end of the measurements.

(v) Plot the forward bias V-I characteristics of the photocell on a graph paper. Choose the origin at the center of the graph paper. Use only the upper half of the graph paper.

**2. Reverse Bias characteristics of the photocell**

(a) Reverse the connections on the photocell (Refer to the circuit diagram (b)).

(b) Connect the digital milli-voltmeter at the regulated voltage source instead of voltmeter.

(c) Increase the intensity of the sodium light illuminating the photocell by decreasing the distance of the photocell from the source window. Set the position of the photocell so that the deflection of spot is about 5 cm on the scale.

(d) Increase the voltage by rotating the potentiometer (fine) clockwise. Note 10 readings of the reverse voltage ($V_{RB}$) and the corresponding deflection (φ) of the spot on the scale. Follow the table 2 to choose the voltage intervals. Calculate the current ($I_{RB}$) through the photocell.

(e) Ensure the minimum position of the potentiometer at the end of the measurements.

(f) Plot the reverse bias characteristics of the photocell on the lower half of the same graph paper. Choose the origin at the bottom middle center of the graph.

(g) From the reverse bias V-I characteristics, note down the voltage ($V_s$) at which the reverse current becomes zero.

(h) Calculate the workfunction of the cathode material of the given photocell using the formula.

**Observation Tables**:

**1. Calibration of the photocell current with the deflection of the galvanometer**

1 cm = ………nA

**2. Forward Bias characteristics of the photocell**

| S. No. | $V_{FB}$ (V) | φ (cm) | $I_{FB}$ (nA) |
|---|---|---|---|
| 1 | 7 | | |
| 2 | 6 | | |
| 3 | 5 | | |
| 4 | 4 | | |
| 5 | 3 | | |
| 6 | 2 | | |
| 7 | 1 | | |
| 8 | 0.8 | | |
| 9 | 0.6 | | |
| 10 | 0.4 | | |
| 11 | 0.2 | | |
| 12 | 0 | | |

# Important Note:

**Maximum deflection should be restricted to 30 cm for the safe operation of the galvanometer.**

**3. Reverse Bias characteristics of the Photocell**

| S. No. | $V_{RB}$ (mV) | φ (cm) | $I_{RB}$ (nA) |
|---|---|---|---|
| 1 | 0 | | |
| 2 | 50 | | |
| 3 | 100 | | |
| 4 | 150 | | |
| 5 | 200 | | |
| 6 | 250 | | |
| 7 | 300 | | |
| 8 | 350 | | |
| 9 | 400 | | |
| 10 | 450 | | |

**Calculations:**

1. To find the accuracy of the digital milli-voltmeter:

Complete accuracy specification: ±(% reading + number of LSD),

where,

Reading = true value of the signal measured by the Digital multimeter,

LSD = Least significant digit

Given, accuracy specification = ±(0.5% + 3),

Range of the Digital milli-voltmeter used, X.XXX

Accuracy of full range of 2 V = 1.999 ± [(1.999)(0.5)/100 + 0.003] volts

2. To find the work function of the cathode material

Given, $h = 6.63 \times 10^{-34}$ J/s, $c = 3 \times 10^{8}$, λ = 5890 Å.

Using, $V_s$ = …….mV

$\Phi = hc/\lambda - eV_s$

= …… (eV) - …... (eV)

= …………….Electron-Volts (eV)

**Results:**

The forward and reverse V-I characteristics of the photocell were studied. The work function of the cathode material of the photocell was found to be …..eV. The accuracy of the digital panel meter which measures the stopping voltage was…….

**Conclusions**:

The photocell when illuminated with the sodium light was found to produce electric current in nano-amperes. This demonstrates the photoelectric effect using the photocell. The work function the cathode material of the photocell was found using the photoelectric effect.

**Precautions**:

1) Ensure the zero position of the potentiometer before switching on the power supply to the kit.

2) Ensure the connections before the start of the measurements as well as throughout the experiment. Ensure the reading of the zero voltage in each set of measurements.

3) The illumination on the photocell should be carefully adjusted in order to produce deflection of the galvanometer within the range of the given scale.

4) If the initial reading in the voltmeter is not zero for the minimum position of the potentiometer, note down the readings as displayed. The calculations are to be performed without applying the zero error correction. The reverse bias curve on the graph paper can be extrapolated to the V = 0 axis with guide to eye.

5) Ensure that the BG is tuned and the reflection from the mirror on the graduated scale is already set. In case the reflected spot is not seen on the scale, take the help of the instructor.

6) In the reverse bias mode the last observation should be taken for the zero deflection (i.e., zero current) and the corresponding value of the stopping voltage should be noted down.

7) The distances and the suggested values of the voltage in the observation tables are for the reference. Use any nearby values during the experiment and note down the readings as seen in the measuring instruments.

8) Connecting wires should be removed carefully and tied after the finish of the experiment.

**Experiment No 2**

## Study of the Hall Effect

**Aim**:

1. Calibrate an electromagnet: To measure the Magnetic field versus the supply current through the magnetic coils

2. To measure the Hall voltage versus the magnetic field at the constant sample current

3. To measure the Hall voltage versus the sample current at the constant magnetic field

4. To determine the Hall coefficient and the carrier concentration

**Apparatus**: Electromagnet, Current supply (in Amperes) for the electromagnet, Gauss meter, Hall probe, Constant current source (in milli-amperes), Digital Millivoltmeter

**Circuit Diagram:**

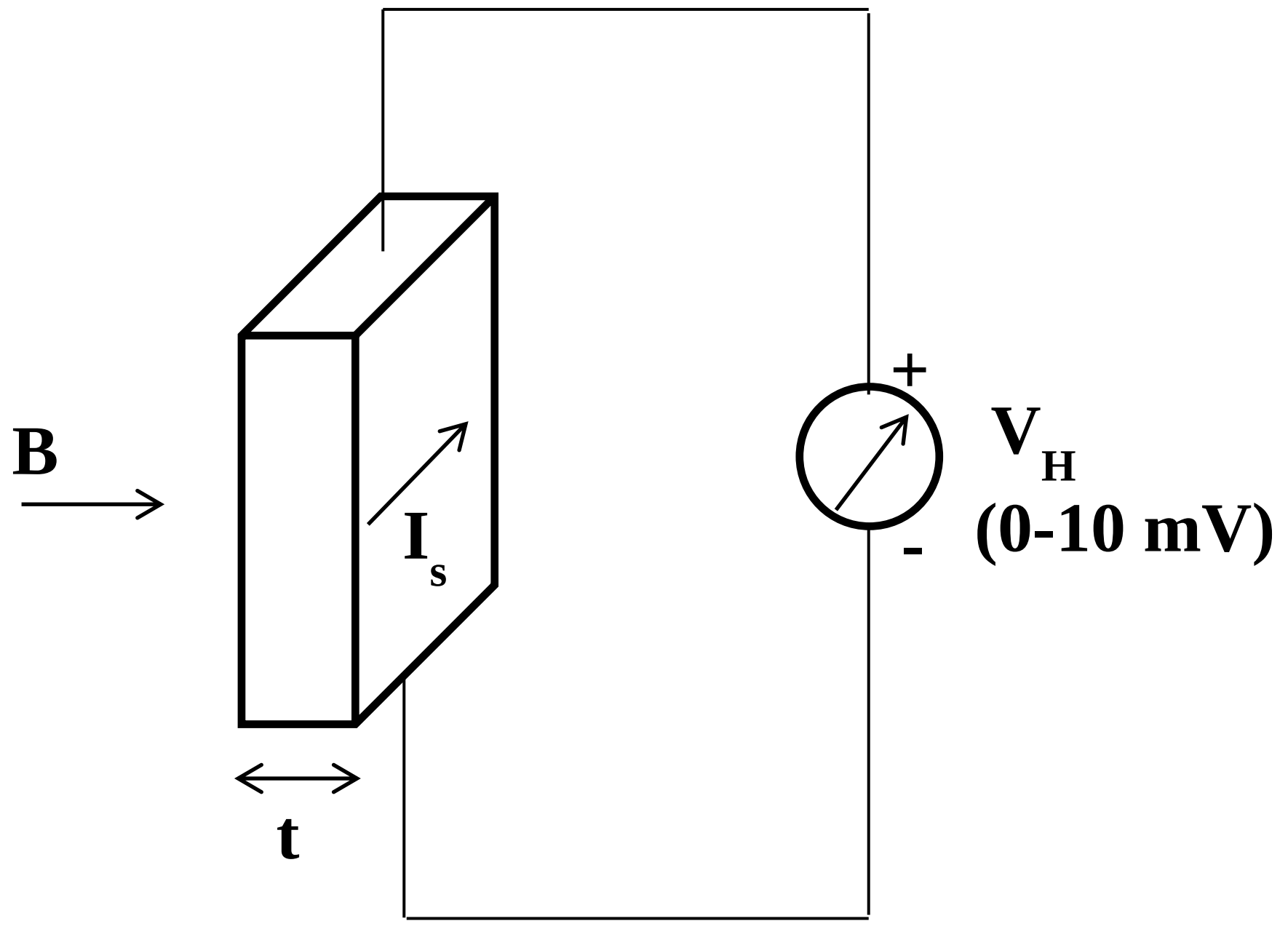

**Formula**: 1. $V_H = (1/ne)\, I_sB/t$ mV

2. $R_H = 1/ne = V_Ht/(I_sB)$ $m^3/C$

3. $n = 1/(eR_H)$ $1/m^3$

Where, $V_H$ = Hall voltage

t = Thickness of the sample

I = The current through the sample

B = The applied magnetic field

n = Carrier concentration

e = Electronic charge

**Procedure**:

**1. Calibration of the electromagnet**

(a) Measure the gap between the pole pieces of the electromagnet and note down in the journal.

(b) Locate the Hall probe in the digital Gauss meter. Adjust the zero reading in the Gauss meter when the Hall probe is away from the magnetic field. The minimum reading should be noted down without zero correction.

(c) Ensure the zero position of the potentiometer on the Current source to the electromagnet. Switch ON the power supply. Insert the Hall probe between the pole pieces and adjust it in the center of the pole pieces of the electromagnet.

(d) Increase the source current in the steps of 0.5 Amps and note the corresponding magnetic field as shown in the Gauss meter. Use the table 1 for the steps to be followed.

(e) Reduce the source current to zero by using the potentiometer on the current source.

(f) Remove the Hall probe of the Gauss meter from the space between the electromagnet.

(g) Plot a graph between the source current versus the magnetic field. Choose the origin at the bottom left corner of the graph paper. Draw a straight line which is the best fit and has maximum number of data points instead of joining the data points one by one.

**2. Hall voltage versus magnetic field**

(a) Insert the sample in the space between the electromagnet and adjust it in the center. Connect the circuit as shown in the figure.

(b) Increase the current through the sample using the potentiometer on the sample current source to 2 mA and keep it constant.

(c) Increase the source current to the electromagnet and measure the voltage in milli-voltmeter. Follow table 2 for the suggested readings.

(d) Reduce the source current to zero by using the potentiometer.

(e) Increase the constant current through the sample to 4 mA and repeat the step (c).

(f) Note down the magnetic field using the calibration curve in part 1 and write in the observation table.

(g) Plot a graph between the Hall voltage versus the magnetic field for constant sample current. Choose the center at the bottom left corner. Find the slope of each curve in the graph and calculate the Hall coefficient using the formula.

(h) Find the mean value of the Hall coefficient and calculate the carrier concentration using the formula (3).

**3. Hall voltage versus sample current**

(a) Reduce the potentiometers on the sample current source and the electromagnet current sources to zero.

(b) Increase the current through the electromagnet using the potentiometer on the constant current source to 1 Amps and keep it constant.

(c) Increase the sample current and measure the Hall voltage in millivoltmeter. Note 5 readings. Follow table 3 for the steps.

(d) Reduce the sample current to zero by using the potentiometer.

(e) Increase the constant current through the electromagnet to 2 Amps and repeat the step (c).

(f) Note down the magnetic field using the calibration curve in part $I_s$ in the observation table.

(g) Plot a graph between the Hall voltage versus the sample current for constant magnetic field. Choose the center at the bottom left corner. Find the slope of each curve in the graph and calculate the Hall coefficient using the formula.

(h) Find the mean value of the Hall coefficient and calculate the carrier concentration using the formula (3).

**Observation Table**:

**1. Calibration of the electromagnet**

| S. No. | Source Current (amps) | Magnetic Field (Gauss) | Magnetic Field ($Wb/m^2$) |
|---|---|---|---|
| 1 | 0 | | |
| 2 | 0.5 | | |
| 3 | 1 | | |
| 4 | 1.5 | | |
| 5 | 2 | | |
| 6 | 2.5 | | |
| 7 | 3 | | |
| 8 | 3.5 | | |

**2. Hall voltage versus Magnetic field**

| S. No | Source Current (amps) | Magnetic Field (Gauss) | Magnetic Field ($Wb/m^2$) | $I_S$ =2 mA | $I_S$ = 4 mA | $I_S$ = 6 mA |
|---|---|---|---|---|---|---|
| | | | | $V_H$ (mV) | $V_H$ (mV) | $V_H$ (mV) |
| 1 | 0 | | | | | |
| 2 | 0.5 | | | | | |
| 3 | 1 | | | | | |
| 4 | 1.5 | | | | | |
| 5 | 2 | | | | | |
| 6 | 2.5 | | | | | |
| 7 | 3 | | | | | |
| 8 | 3.5 | | | | | |

**3. Hall voltage versus sample current**

| S. No | Sample Current ($I_s$) (mA) | I = 1 A<br>B = …..Wb/m$^2$ | I = 2 A<br>B =……Wb/m$^2$ | I = 3 A<br>B =……Wb/m$^2$ |
|---|---|---|---|---|
| | | $V_H$ (mV) | $V_H$ (mV) | $V_H$ (mV) |
| 1 | 0 | | | |
| 2 | 1 | | | |
| 3 | 2 | | | |
| 4 | 3 | | | |
| 5 | 4 | | | |

**Calculations**:

Given: charge of electron, e = 1.6 x 10 $^{-19}$ Coulombs,

Distance between the pole pieces of the electromagnet, D = …..cm

Thickness of the sample, t = 0.3 cm

1. Determination of Hall Coefficience, $R_H$

(a) From the graph, $V_H$ versus B,

$I_s$ = 2 mA, $R_H = m_1$ x $t/I_s$ = ….. m$^3$/C using (on graph) $m_1 = (V_{H2}-V_{H1})/(B_2-B_1)$ = …….,

$I_s$ = 4 mA, $R_H = m_2$ x $t/I_s$ = ….. m$^3$/C using (on graph) $m_2 = (V_{H2}-V_{H1})/(B_2-B_1)$ = ……..,

$I_s$ = 6 mA, $R_H = m_3$ x $t/I_s$ = ….. m$^3$/C using (on graph) $m_3 = (V_{H2}-V_{H1})/(B_2-B_1)$ = ……..,

Mean $R_H$ = (……+……+…..)/3

(b) From the graph, $V_H$ versus $I_s$,

B = ……Wb/m$^2$, $R_H = m_1$ x t/B = ….. m$^3$/C using (on graph) $m_1 = (V_{H2}-V_{H1})/(I_{s2}-I_{s1})$ = …….,

B = ……Wb/m$^2$, $R_H = m_2$ x t/B = ….. m$^3$/C using (on graph) $m_2 = (V_{H2}-V_{H1})/(I_{s2}-I_{s1})$ = …….,

B = ……Wb/m$^2$, $R_H = m_3$ x t/B = ….. m$^3$/C using (on graph) $m_3 = (V_{H2}-V_{H1})/(I_{s2}-I_{s1})$ = …….,

Mean $R_H$ = (……+……+…..)/3

2. Determination of carrier concentration, n

(a) From the graph of $V_H$ versus B, $R_H$ =…….. $m^3/C$

$n = 1/(eR_H)$ =……..   $1/m^3$

(b) From the graph of $V_H$ versus $I_s$, $R_H$ =…….. $m^3/C$

$n = 1/(eR_H)$ =……..   $1/m^3$

**Result**:

| Parameters | From $V_H$ vs B studies | From $V_H$ vs $I_s$ studies |
|---|---|---|
| Hall Coefficient, $R_H$ ($m^3/C$) | | |
| Carrier concentration, n ($1/m^3$) | | |

**Conclusions**:

The electromagnet was calibrated using the Gauss meter. The Hall Effect was studied and it was found that the Hall voltage depends on the sample current as well as the magnetic field. The Hall coefficient was found from both $V_H$ vs B and $V_H$ vs I and the carrier concentration at room temperature was calculated for the given semiconducting material.

**Precautions**:

1) Ensure the zero position of the potentiometer before switching on the power supply to the electromagnets.

2) Ensure the connections, and the appropriate voltage and current ranges in the meters before the start of the measurements.

3) Ensure the reading of the zero voltage in each set of measurements.

4) Occupy more than 50% of the graph paper to ensure good presentation of the results.

5) Ensure that the position of the sample remains the same between the pole pieces throughout the experiment.

6) Readings should be taken while increasing the magnetic field. If the suggested reading is exceeded appreciably while increasing the field then the next closest reading may be recorded instead of reducing the field.

7) In $V_H$ vs B measurement, the field must be reduced to zero before increasing the sample current $I_s$ for each set of readings.

8) Connecting wires should be removed carefully and tied after the finish of the experiment.

**Experiment No 3**

## Determination of e/m by Thomson's method

**Aim**: To determine the charge to mass ratio (e/m) for an electron by Thomson's method

**Apparatus**: Cathode ray tube (CRT), Power supply for CRT, Wooden frame, Deflection magnetometer, Wooden bench, Bar magnets, Stop watch

**Circuit Diagram:**

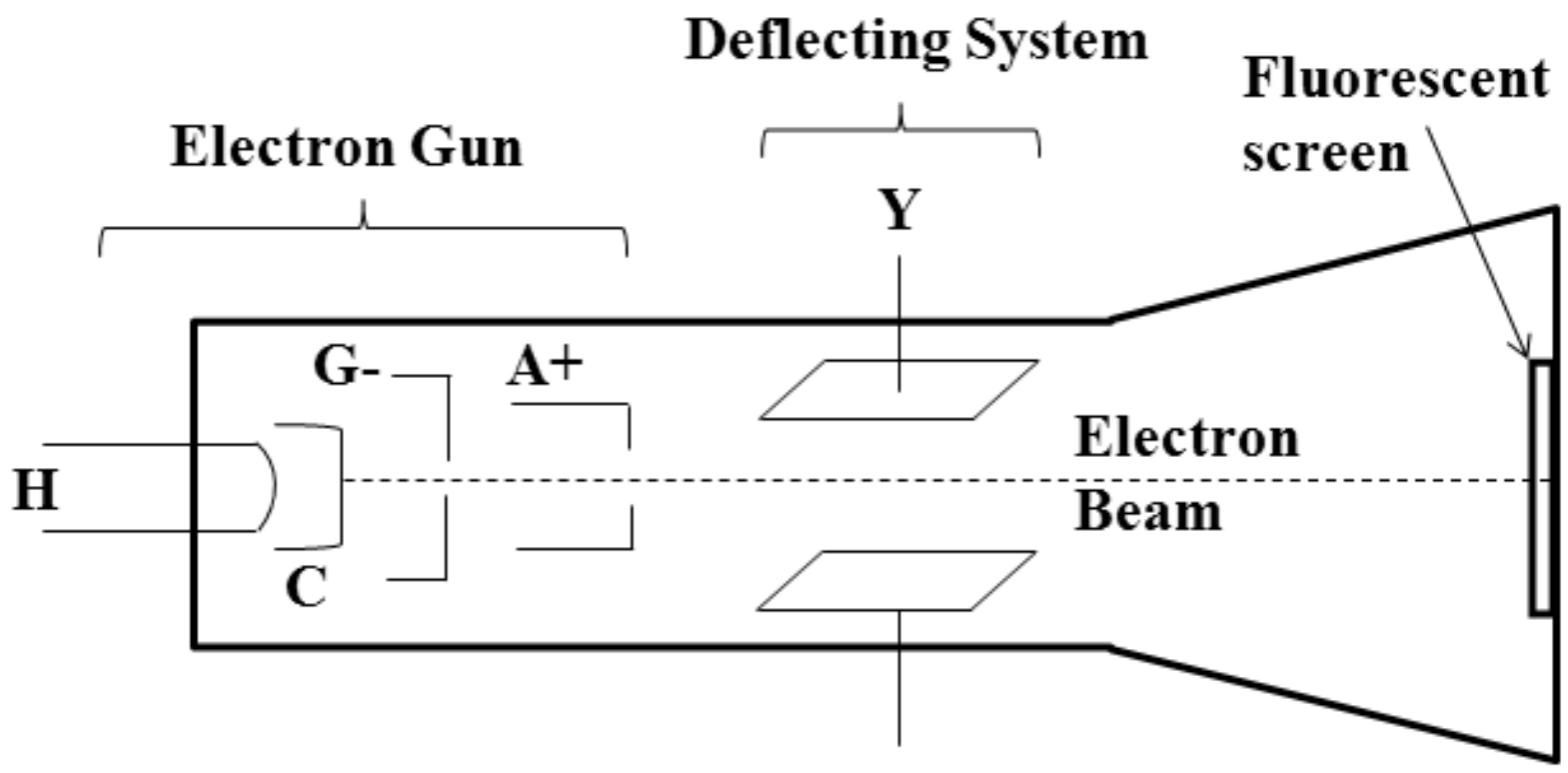

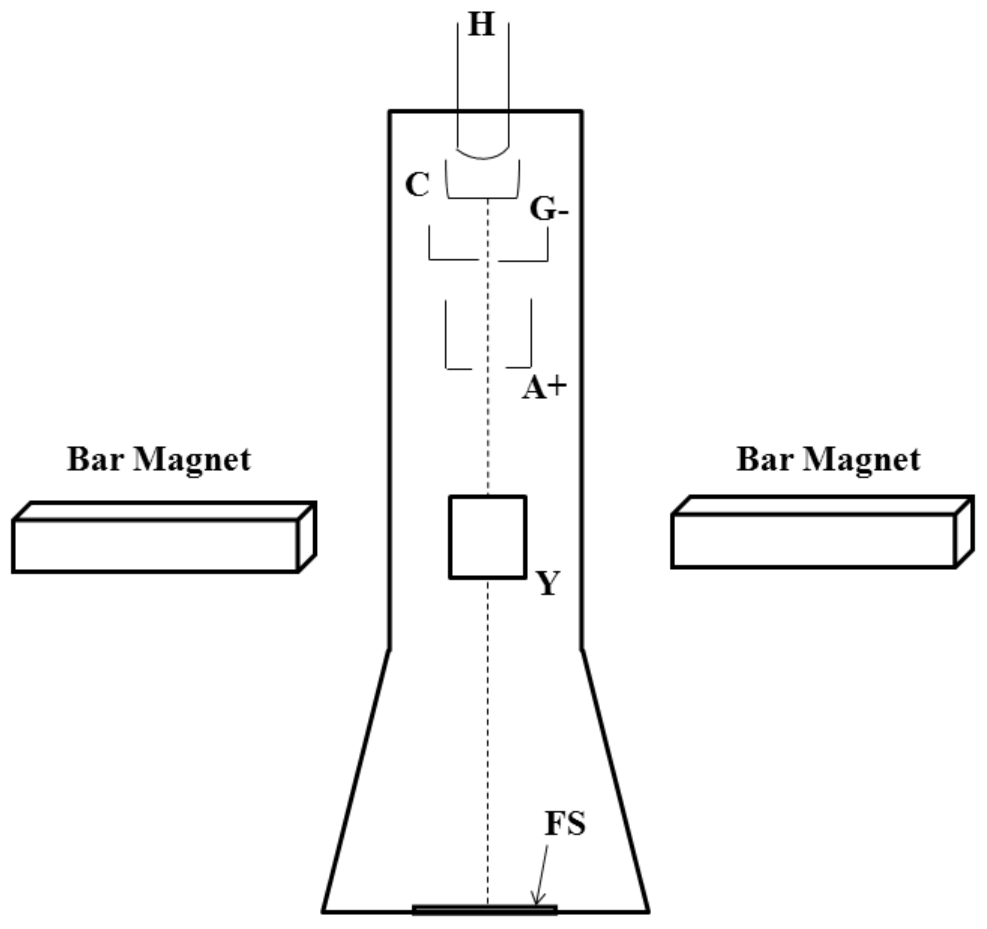

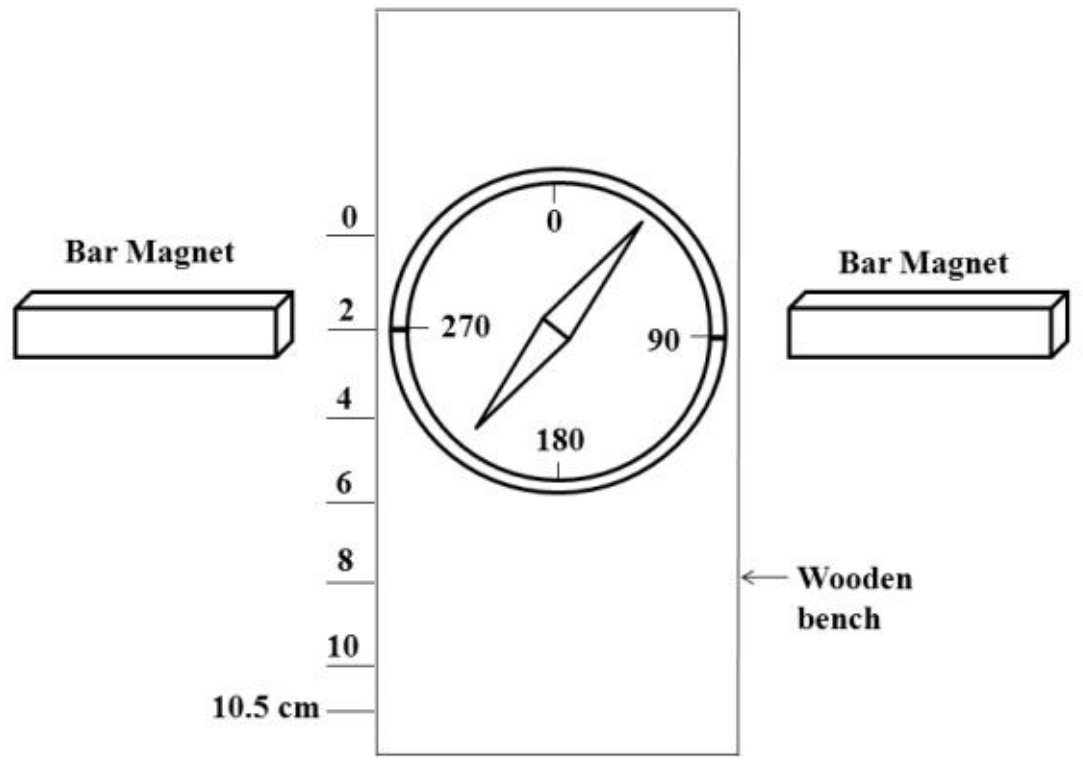

**Formula**: 1. $e/m = y\,(V/d)\,p(l + p/2)/A^2$ C/kg

2. $B = B_0\,(T_0^2/T^2)\sin\varphi$ Wb/m$^2$

Where, e = charge of an electron,

m = mass of an electron

V = voltage applied to deflection plates

Y = displacement of the spot

d = separation between the deflecting plates

p = length of deflecting plates

*l* = distance between end of deflecting plates and screen

B = magnetic induction

$B_0$ = horizontal component of earth's magnetic field

T = time period of oscillation of deflection magnetometer needle in the presence of two bar magnets

$T_0$ = period of oscillation of deflection magnetometer needle in the presence of earth's magnetic field

φ = inclination of the magnetic needle to the earths' magnetic field

A = area under the curve plotted between x and [B(L-x)]

L = distance of electron gun exit to the screen

**Procedure**:

**1. Positioning the wooden bench**

(a) Align the wooden bench perpendicular to the magnetic meridian using the given deflection magnetometer. For this step, fix the wooden frame and bench together.

(b) Place the deflection magnetometer on the bench such that the needle aligns with the 0-0 degree marks of the deflection which is also parallel to the length of the wooden bench.

(c) Mark the position of the wooden bench and it should remain same during the experiment.

**2. Determination of Electric field**

(a) Remove the deflection magnetometer and the wooden frame from the bench.

(b) Place the Cathode ray tube (CRT) in the middle of the wooden bench and switch on its power supply after ensuring the DIF. INT knob in the minimum position.

(c) Observe the spot on the fluorescent screen of the CRT and if required increase the intensity and focusing of the electron beam. In the absence of electric and magnetic field, the electron beam is not deviated and the spot should remain at the zero reading of the CRT graphical screen.

(d) Increase the DIF INT potentiometer in order to apply voltage to the vertical deflecting plates. This produces electric field and it deflects the electron beam vertically upwards. Note the least count of the voltmeter given on the CRT power supply.

(e) Note down the voltage corresponding to the deflection of 1 cm for the spot on the CRT screen. Note both the values in the observations in the journal.

**3. Balancing configuration with the magnetic field**

(a) In order to balance the electric field with the magnetic field, place the two bar magnets on the wooden bench on the left and the right side of the CRT screen. Refer figure (b) for the placing of the bar magnets on the wooden bench in a symmetric manner in order to produce a perpendicular magnetic field with respect to the electric field and the direction of electron beam (along the length of the Cathode ray tube).

(b) Arrange the distance between the magnets so that the bright spot on the screen returns to the zero position on the CRT graphical screen. This will balance the electric field on the electron beam with the magnetic field of the bar magnets and the given formula is applicable to determine e/m value. The position of the bar magnets should remain same during the experiment.

**4. Measuring the magnetic field**

(a) Reduce the potentiometer DIF INT on the CRT power supply to minimum position. Switch off the power supply and remove the CRT from the wooden bench.

(b) Place the deflection magnetometer on the wooden frame and insert this assembly in the middle of the wooden bench to measure the magnetic field along the length of the electron beam in the Cathode ray tube (i.e., the distance of L = 10.5 cm from the electron gun at the back upto the fluorescent screen). The 0-0 reading the deflection magnetometer should align with the wooden bench. The starting position of the deflection magnetometer should be corresponding to x = 0 cm.

(c) Note down the deflection of both the pointers in the magnetometer and write it in the journal.

(d) Measure the time for 5 oscillations of the pointer in the deflection magnetometer. For setting the oscillations use the additional bar magnet and the stop watch. Reset the stop watch to zero reading and bring the bar magnet close to the magnetometer. In case 5 oscillations are not achieved, use 2, 3 or 4 oscillations for measuring the time period by carefully noting down each time the number of oscillations in the observation table.

(e) Shift the deflection magnetometer to the next x position of 2 cm along the wooden frame and repeat the steps from (c) and (d).

**5. Measurement of time period in the Earth's magnetic field**

(a) Remove the bar magnets from the wooden bench.

(b) Place the deflection magnetometer again at x = 0 cm with 0-0 reading coinciding with the length of the wooden bench.

(c) Reset the stop watch and set up oscillations in the magnetometer using any of the bar magnet.

(d) Measure the time of oscillations using the stop watch. The measurement should be done twice for the number of oscillations which can be between 2 to 5. Measurement of time period only for 1 oscillation should be avoided.

**6. Calculation of e/m value**

(a) Calculate the magnetic field, B using the formula 2 and note down in the observation table.

(b) Plot a graph between (L-*x*) B vs *x* by choosing origin at the left corner of the graph paper. Refer the Appendix for the choice of the origin and the scale. Count the number of 1 $cm^2$ and 1 $mm^2$ boxes under the curve and note on the graph paper.

(c) Calculate the area under the curve by multiplying the total area (in cm) by the appropriate units of the x and y axes.

(d) Calculate the e/m value using the formula 1 and compare it with the theoretically calculate value of e/m.

**Observation Table**:

y = ……cm, V = ….volts, $B_0$ = ….Wb/$m^2$, L = 10.5 cm

Time of 5 oscillations in Earth's magnetic field, $t_0$ = ….sec

Time period of 1 oscillation in Earth's magnetic field, $T_0 = t_0/5$ = ….sec

| S. No. | x (m) | $\Phi$ | | | Time of 5 oscillations t (sec) | | | Time period T = t/5 | Magnetic Field B (wb/$m^2$) = $B_0$ ($T_0^2/T^2$) sin($\Phi_m$) | (L-x) (m) | B(L-x) (wb/m) |
|---|---|---|---|---|---|---|---|---|---|---|---|
| | | $\Phi_1$ | $\Phi_2$ | $\Phi_m$ | $t_1$ | $t_2$ | $t_m$ | | | | |
| 1 | 0 | | | | | | | | | | |
| 2 | $2 \times 10^{-2}$ | | | | | | | | | | |
| 3 | $4 \times 10^{-2}$ | | | | | | | | | | |
| 4 | $6 \times 10^{-2}$ | | | | | | | | | | |
| 5 | $8 \times 10^{-2}$ | | | | | | | | | | |
| 6 | $10 \times 10^{-2}$ | | | | | | | | | | |
| 7 | $10.5 \times 10^{-2}$ | | | | | | | | | | |

**Calculations**:

Given: p = 2 cm, *l* = 1.82 cm, d = 0.3 cm, L = 10.5 cm

Using, y = 1 cm, V = ….volts

From the graph between x versus B(L-x), area under the curve, A =….. Wb

$$e/m = y\ (V/d)\ p(l + p/2)/A^2$$

= …….. Coulombs/Kg

Theoretical value of e/m = ………Coulombs/Kg

**Results**:

The e/m of an electron using the Thomson's method was ……. Coulombs/Kg.

**Conclusions:**

The Thomson's method to determine the e/m of electrons was studied. The ratio of e/m was found to be…..which compares well with the theoretically estimated value of the e/m. The e/m was found in the Thomson's experiment without requiring the value of e or m of the charge particles. If the mass of the charge particle increases with the velocity approaching to the relativistic values, the e/m will decrease.

**Precautions**:

1) Magnetic meridian should be found by keeping the magnets away from the magnetic needle.

2) The wooden bench should remain in the magnetic meridian throughout the experiment.

3) Stopwatch readings should be taken accurately.

4) $t_0$ in the Earth's magnetic field should be noted for two sets of oscillations.

5) The number of oscillations is a suggested reading. If the deflection magnetometer is not sensitive for 5 oscillations, only 2 or 3 oscillations may be taken to calculate the time period (T).

**Experiment No 4**

## Study of cathode ray oscilloscope

**Aim**:

1. To use cathode ray oscilloscope (CRO) to determine the amplitude and the frequency of the sinusoidal voltage.

2. To determine the phase difference between the two sinusoidal electrical signals using cathode ray oscilloscope

**Apparatus**: Cathode ray oscilloscope, Ac signal generator, Transformer, Ac voltmeter, RC circuit, Connecting wires

**Figures:**

**a. Amplitude and frequency**

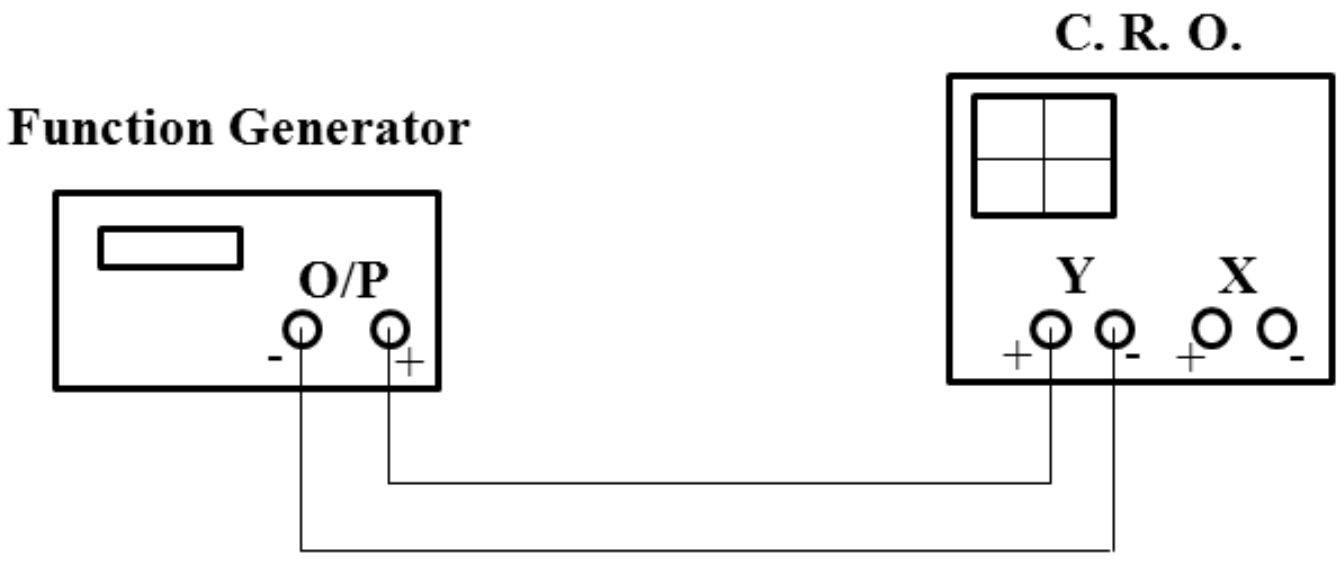

**b. Phase determination**

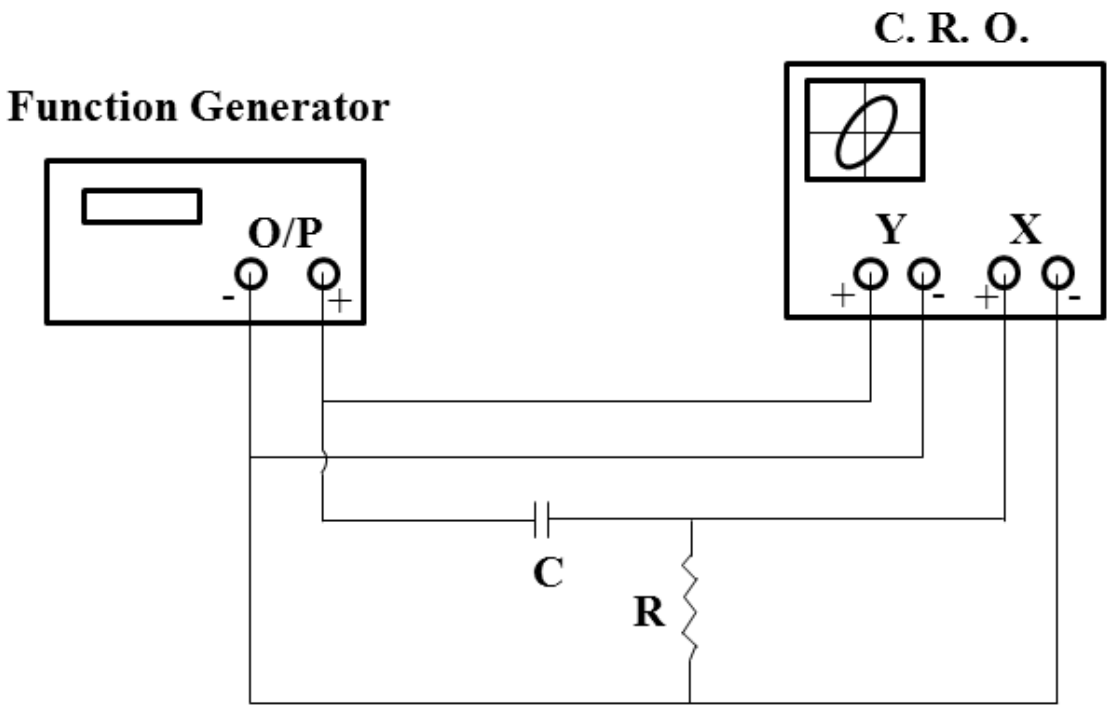

**c. Determination of unknown frequency**

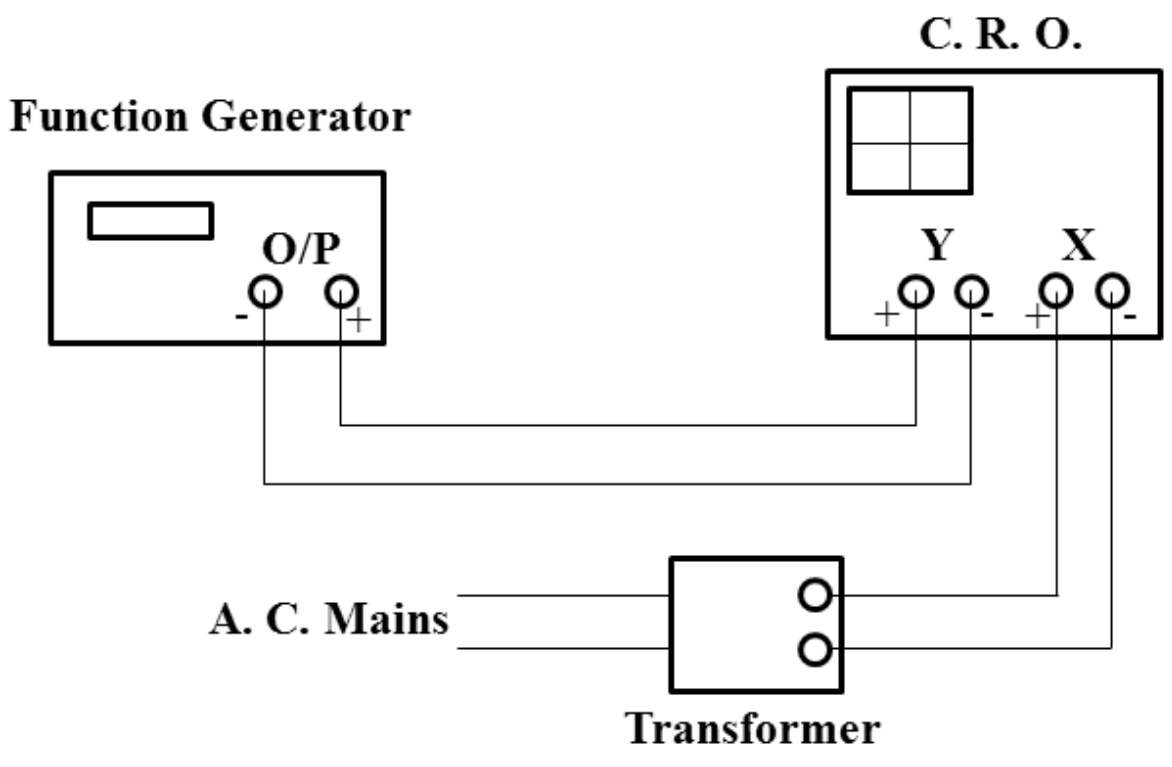

**Formulae:**

1. Measurement of ac voltage

$$V_{rms} = a/(2)^{1/2}, \quad a = (n/2)\ p$$

where, $V_{rms}$ = room mean square of the ac voltage

a = amplitude of sinusoidal voltage

n = vertical length of display (peak to peak) in number of divisions

p = reading of corresponding volt/division knob

2. Calibration of frequency of oscillator,

$$T = d \times t \qquad \text{and} \qquad f = 1/T$$

where, T = time period

d = spread of sine wave

t = time base setting (time/div)

f = frequency of the oscillator

3. Measurement of frequency

$$f_y = [T_x/T_y]\ f_x \quad \text{Hz}$$

where, $T_x$ = Number of tangency point along x-axis

$T_x$ = Number of tangency point along y-axis

$f_x$ = Frequency of signal applied to x-input

$f_y$ = Number of signal applied to y-input

4. CRO in x-y mode, $\alpha_p = \sin^{-1}[y_1/y_2]$

where, $\alpha_p$ = experimental value of the phase difference

$y_1$ = intercept of ellipse on the y-axis

$y_2$ = maximum vertical deflection

$$\alpha_t = \tan^{-1}[1/(2\pi\ f\ RC)]$$

where, $\alpha_t$ = theoretical value of the phase difference

f = frequency

R = resistance and C = capacitance

**Procedure**:

**1. Switch On the cathode ray oscilloscope. Adjust the intensity and the focus of the bright spots on the CRO screen.**

**2. Using the x-position and y-position knobs, center the bright spot.**

**3. Measurement of amplitude and frequency of the signal using CRO**

(a) Set the function generator for a frequency of 1 kHz by selecting the appropriate range and the position of the coarse and fine knobs. The signal voltage can be controlled using the LEVEL knob on the function generator and pressing left corner button to the $V_{p\text{-}p}$ instead of Frequency. Ensure the sine waveform button is pressed. All other buttons should remain unpressed on the function generator.

(b) Use the circuit diagram (a) to connect the function generator to the channel CH1 (also called Y input) of the CRO. Choose the Mono mode (corresponds to the unpressed button position) and the CH1 (unpressed button position).

(c) Adjust the voltage control knob to get the waveform in the measurable range on the screen of the CRO.

(d) Use the time base (t/div) knob to get a stable sinusoidal waveform on the CRO screen.

(e) Note the vertical divisions on the CRO screen for the peak to peak distance of the waveform.

(f) Note the corresponding volt/div from the knob in the observation table 1.

(g) Calculate the amplitude of the input signal using div x (volt/div).

(h) Repeat the steps (c) to (g) for three different values of the input signal.

(i) Change the function generator panel display to the FREQ mode by unpressing the left corner button.

(j) Note the frequency from the digital panel in the observation table 2.

(k) Note the number of divisions for 1 wavelength of the waveform on the CRO screen. Use the time axis of the signal on the screen.

(l) Note down the time base (t/div) knob position in the observation table 2.

(m) Calculate the frequency of the signal on the CH1 by the formula 2.

(n) Repeat the steps (j) to (m) for three different values of the signal frequencies.

**4. Measurement of the phase difference in the dual mode of the CRO**

(a) Connect the circuit as shown in figure (b). In this case the signal from the function generator is connected directly to the CH1 as in step 3 and the same signal is connected at the input of the RC circuit. The signal at the output of the RC circuit (with a phase difference) is connected to the CH2 (also called X-input of CRO).

(b) Press the mono/dual press button to switch the CRO in the dual mode. In this mode both the signals, on CH1 and CH2, will be displayed on the CRO screen.

(c) Adjust the volt/div knob for both the signals and use the x-position and time/base knob to display the signals in the suitable voltage and time scales. One of the signals should cross the zero position on the graphical screen of the CRO.

(d) Note down the time period and the time difference between the signals in the observation table 3.

(e) Using the formula, calculate the phase difference between the signals. Compare the phase difference with the theoretical value by noting down the resistance, capacitance of the RC circuit and the frequency of the signal estimated from the time period.

**5. Measurement of phase difference in x-y mode of the CRO**

(a) Unpress the dual mode button and press the x-y push button on the CRO front panel.

(b) Adjust the x and y-position knobs and place the ellipse symmetric around the voltage and time axes. Adjust the volts/div knob if the size of the ellipse needs to be changed to cover the maximum portion of the CRO graphical screen.

(c) Note the readings of the vertical intercept, $y_1$ and the horizontal tangent, $y_2$ for the ellipse on the screen (Refer to the Appendix). Use the observation table 3.

(d) Calculate the phase difference using the formula and compare with the theoretical value.

**6. Measurement of unknown frequency of the signal using CRO**

(a) Connect the circuit as shown in figure (c) and switch ON the given ac transformer. In this case the signal from the function generator is applied at the CH1 of the CRO and the unknown signal is applied on the CH2.

(b) Set the frequency of the signal close to 100 Hz using the coarse knob on the function generator. Fine tune the frequency (using the Fine knob) till a stable Lissajous pattern (as a horizontal figure of 8) is seen on the CRO screen. Refer to the Appendix for Lissajous patterns.

(c) Note the number of horizontal tangent points ($T_x$) and vertical tangent points ($T_y$) of the Lissajous pattern.

(d) Note the known frequency of the signal of the function generator in the observation table 4.

(e) Using the given formula, calculate the unknown frequency of the input signal on CH2.

(f) Repeat the steps (b)-(e) for three additional frequencies in the multiple of 50 Hz and in each case determine the unknown frequency (of ac mains) by obtaining a near stable Lissajous pattern on the screen of the CRO.

**Observation Tables**:

**1) Measurement of the amplitude of the input signal**

| Sr. No. | n (peak to peak divisions) | p (volt/div) | a = p (n/2) (volts) | $V_{rms}$ (volts) = $a/(2)^{1/2}$ |
|---|---|---|---|---|
| 1 | | | | |
| 2 | | | | |
| 3 | | | | |
| 4 | | | | |

**2) Measurement of frequency**

| Sr. No. | Frequency selected in function generator (kHz) | Divisions on the CRO screen for one period of sine wave, d | Time base setting (time/div)<br><br>t | Time Period of the input signal<br><br>T = d x t | Frequency of the input signal<br><br>f = 1/T (kHz) |
|---|---|---|---|---|---|
| 1 | 1 | | | | |
| 2 | 1.5 | | | | |
| 3 | 2 | | | | |
| 4 | 2.5 | | | | |

**3) Measurement of the phase angle between two signal waveforms using CRO for f = …..Hz**

| Sr. No. | R (kΩ) | C (µF) | $\alpha_T$ (degree) | Dual mode of the CRO | | | x-y mode of the CRO | | |
|---|---|---|---|---|---|---|---|---|---|
| | | | | Δt (msec) | T (msec) | $\alpha_p = [\Delta t/T]360^o$ | $y_1$ | $y_2$ | $\alpha_p = \sin^{-1}[y_1/y_2]$ |
| 1 | | | | | | | | | |

**4) Measurement of unknown frequency of a.c. mains**

| Sr. No. | Frequency of signal generator $f_y$ (Hz) | Number of tangent points | | $f_x$ (Hz) = $(T_y/T_x) \times f_y$ Hz |
|---|---|---|---|---|
| | | $T_x$ | $T_y$ | |
| 1 | 50 | | | |
| 2 | 100 | | | |
| 3 | 150 | | | |
| 4 | 200 | | | |

**Calculations (1 from each observation table)**:

1. $V_{rms} = a/(2)^{1/2}$,   $a = (n/2)\ p$

2. $T = d \times t$  and  $f = 1/T$ Hz

3. $f_x = (T_y/T_x)\ f_y$  Hz

4. $\alpha_p = \sin^{-1}[y_1/y_2]$  and  $\alpha_T = \tan^{-1}[1/(2\pi\ f\ RC)]$

**Result**: The frequency of the unknown source was found to be ……Hz. The phase difference between the two electrical signals was found to be…..and ……degrees. The theoretically estimated value was….degrees.

**Conclusions**: The CRO was used to find out the amplitude of the input signal and the unknown frequency of the input signal using the formation of the Lissajous pattern. The CRO was used to find out the phase difference between the two electrical signals generated using an R-C circuit. The theoretical value was estimated for the R-C generator, and compared with the experimentally found values using the x-y and dual modes of the CRO operation.

**Precautions**:

1) The intensity of the screen is kept at the minimum, and reduced to zero if not in use.

2) The Lissajous patterns should cover the majority portion of the CRO screen for improving the precision of the measurements.

3) The waveforms seen on the screen should be placed symmetrically about the time axis.

4) Before obtaining the Lissajous patterns to measure the unknown frequency, both the signals (on CH1 and CH2) should be verified by using the MONO mode of the CRO for Channel 1 and Channel 2. The CH1/2 knob can be used to switch between the channels in the MONO mode, while the x-y mode should be in the unpressed position.

5) Connecting wires should be removed carefully and tied after the finish of the experiment.

**Experiment No 5**

## V-I Characteristics of the diode and determination of the energy gap of the semiconductor

**Aim**:

1. To study the V-I characteristics of the Germanium PN Junction Diode in the Forward Bias mode.

2. To study the V-I characteristics of the Germanium PN Junction Diode in the Reverse Bias mode

3. To determine the energy gap of the semiconductor material of the given PN Junction Diode operating in the Reverse Bias mode.

**Apparatus**: Experimental Kit of Germanium PN Junction diode, Regulated dc power supply, Milli-Ammeters (Range 0 to 50 mA for forward bias current, 0 to 2 mA digital panel meter for the reverse bias current and 0 to 500 µA for the reverse saturation current in the energy gap setup), Milli-Voltmeter (0 to 1000 mV for forward bias voltage), Voltmeter (0 to 20 Volts attached on the dc source for the reverse bias voltage), Water bath, Heater, Thermometer, Connecting wires

**Circuit Diagrams: Refer to the supplementary material**

**(a) Forward bias** **(b) Reverse bias**

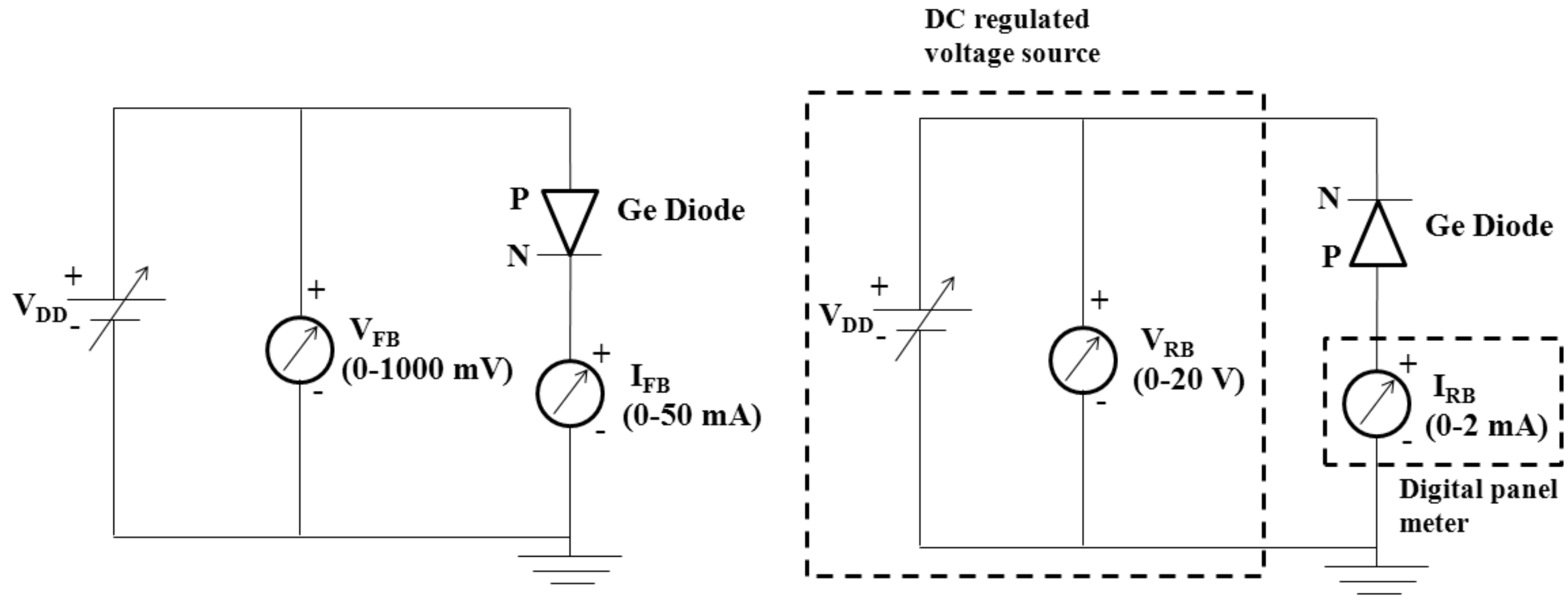

**(c) Energy gap setup**

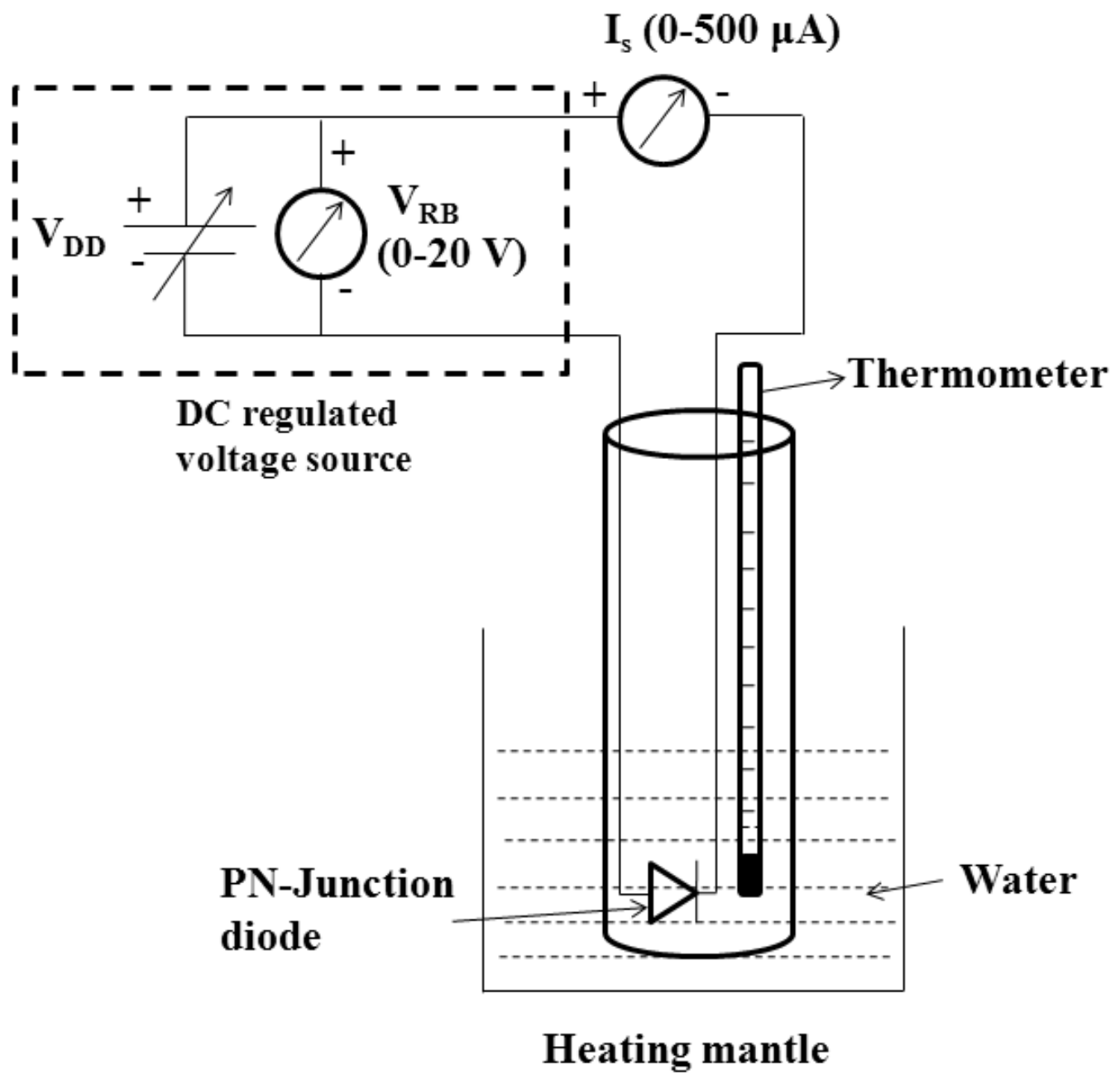

**Formulae:** (1) Static resistance, $R_s = V_{FB}/I_{FB}$ ohms (Ω)

(2) Dynamic resistance, $R_d = (V_{FB2}-V_{FB1})/(I_{FB2}-I_{FB1})$ Ω

(3) Reverse saturation current, $I_s \propto \exp(-E_g/kT)$

Where, k = Boltzmann's constant

T = Temperature in Kelvin

Energy gap, $E_g = |m| \times 1.986 \times 10^{-4}$ eV

m = Slope of $Log_{10}I_s$ versus 1/T

**Procedure**:

**1. Forward Bias characteristics**

(a) Connect the Germanium diode to study the V-I characteristics.

(b) Connect the regulated voltage source supply across the given Germanium diode in series with the milliammeter. Refer to figure (a) for the forward bias configuration.

(c) Ensure the zero position of the potentiometer on the voltage source.

(d) Connect the millivoltmeter in parallel the regulated voltage source as shown in figure (a).

(e) Note the least count of the milliammeter and millivoltmeter for the selected ranges.

(f) Gradually increase the forward voltage across the Germanium PN Junction diode by rotating the potentiometer clockwise. Note 10 readings of the voltage ($V_{FB}$) and the corresponding current ($I_{FB}$) through the diode. Follow the table 1 to choose the voltage intervals.

(g) Ensure the zero position of the potentiometer on the voltage source.

(h) Plot the forward bias V-I characteristics of Germanium diode on a graph paper. Choose the origin at the center of the graph paper. Write the scale of x-axis and y-axis on the top right corner of the graph paper.

(i) Plot the tangent to the forward V-I characteristics of Germanium Diode. Use the linear portion of the curves which is characteristic of the switch-ON state of the diode and the tangent should contain the maximum data points. Find out the intercept on the voltage axis and note it down as the cut-in voltages of the Germanium Diode. Refer to the supplementary material in the manual for the forward characteristics and determination of the cut-in voltage.

(j) Estimate the static forward resistance ($R_S$) of the Germanium Diode by choosing one point ($V_{FB}$, $I_{FB}$) on the V-I characteristics. Use the switch-ON state of the diode in their forward V-I characteristics.

(k) Choose a set of ($V_{FB2}$, $I_{FB2}$) and ($V_{FB1}$, $I_{FB1}$) in the linear region of the V-I characteristics. Estimate the dynamic resistance ($R_d$) of the Germanium Diode from the forward slopes of the V-I characteristics.

**2. Reverse Bias characteristics of the Germanium Diode**

(a) Connect the Germanium diode in the reverse bias mode as shown in figure (b). Reverse the polarity of the diode with respect to the voltage source.

(b) Ensure the zero position of the potentiometer on the voltage source.

(c) Use the voltmeter built-in the voltage source for the reverse bias characteristics. Note the least count of the voltmeter.

(d) Gradually increase the forward voltage across the Germanium PN Junction diode by rotating the potentiometer clockwise. Note up to 10 readings of the voltage ($V_{RB}$) and the corresponding reverse current ($I_{RB}$) through the diode. Follow the table 2 to choose the voltage intervals.

(e) Ensure the zero position of the potentiometer on the voltage source.

(f) Plot the reverse bias V-I characteristics of the Germanium diode along with its forward bias characteristics on the graph paper. Write the scale of x-axis and y-axis on the bottom left corner of the graph paper.

(g) Plot the tangent to the reverse characteristics in the saturation region of the curve. Note the reverse saturation current ($I_s$). Refer to the supplementary material in the manual for the reverse characteristics and determination of the reverse saturation current.

**3. Energy gap determination**

(a) Carefully pour the water in the container and dip the glass tube (with diode and thermometer assembly) inside the container.

(b) Connect the diode terminals as shown in the figure (c) for the reverse bias operation. Apply $V_{RB}$ = 6 volts.

(c) Switch ON the heater after connecting it to the on-board power supply and monitor the temperature in the thermometer.

(d) Increase the temperature upto 85 ºC while taking care of the overshoot.

(e) Switch OFF the heating mantle from the on-board supply and wait till the temperature continues to increase above 90 ºC.

(f) Without removing the glass tube from the water bath, wait for the stable temperature reading in the thermometer as the water starts cooling down.

(g) Note down the current in the digital microammeter corresponding to 90ºC as your first reading.

(h) Read the temperature with every 5 degree fall as the water cools down to 40ºC, and note the corresponding reverse saturation current ($I_s$) through the PN Junction Diode. Follow the table 3 to choose the temperature intervals.

(i) Plot ($\log_{10} I_s$) versus 1/T on a separate graph paper by choosing the origin on the bottom left corner. Note the scale on the x-axis and y-axis on the top right corner of the graph paper.

(j) Draw a line and estimate the slope from the graph. The line should ideally contain equal number of data points on either side of it. Refer to the supplementary material in the manual.

(k) Using the slope, calculate the energy gap of the semiconducting material of the PN-Junction diode with the given formulae.

**Observation Table**:

**Least Counts (L.C.) of the measuring instruments**

(1) Milli-ammeter (Forward Bias) with Range (………..): L.C. = 1 Main scale reading = ……

(2) Milli-voltmeter (Forward Bias) with Range (………..): L.C. = 1 Main scale reading = ……

(3) Micro-ammeter (Energy gap set-up) with Range (……….): L. C. = 1 Main scale reading = …

(4) Voltmeter (Energy gap set-up) with Range (………): L. C. = 1 Main scale reading = ……

**1. Forward Bias characteristics**

| S. No. | Germanium Diode | |
|---|---|---|
| | $V_{FB}$ (mV) | $I_{FB}$ (mA) |
| 1 | 0 | |
| 2 | 50 | |
| 3 | 100 | |
| 4 | 125 | |
| 5 | 150 | |
| 6 | 175 | |
| 7 | 200 | |
| 8 | 250 | |
| 9 | 300 | |
| 10 | 350 | |

**2. Reverse Bias characteristics**

| S. No. | Germanium Diode | |
|---|---|---|
| | $V_{RB}$ (V) | $I_{RB}$ (μA) |
| 1 | 0 | |
| 2 | 0.25 | |
| 3 | 0.5 | |
| 4 | 0.75 | |
| 5 | 1 | |
| 6 | 2 | |
| 7 | 3 | |
| 8 | 4 | |
| 9 | 5 | |

**3. Variation of the reverse saturation current of the PN-Junction diode versus temperature**

**$V_{RB}$ = 6 V**

| S. No | Temperature (T) (ºC) | Current, $I_s$ (μA) | Temperature (T) (K) | 1/T ($K^{-1}$) | $Log_{10} I_s$ |
|---|---|---|---|---|---|
| 1 | 90 | | | | |
| 2 | 85 | | | | |
| 3 | 80 | | | | |
| 4 | 75 | | | | |
| 5 | 70 | | | | |
| 6 | 65 | | | | |
| 7 | 60 | | | | |
| 8 | 55 | | | | |
| 9 | 50 | | | | |
| 10 | 45 | | | | |

**Calculations:**

1. The static resistance, $R_s$ in the linear region of the V-I characteristics point, corresponding to ($V_{FB}$, $I_{FB}$) is,

$R_s = V_{FB}/I_{FB}$ = …../…. = …..Ω

2. The dynamic resistance, $R_d$ in the linear region of the V-I characteristics of the diode

| Diode | $V_{FB1}$ (V) | $I_{FB1}$ (mA) | $V_{FB2}$ (V) | $I_{FB2}$ (mA) | $R_d = (V_{FB2}-V_{FB1})/(I_{FB2}-I_{FB1})$ |
|---|---|---|---|---|---|
| Germanium | | | | | ……………..(Ω) |

3. From the graph of $Log_{10}I_s$ versus 1/T, Slope = …………

Substituting in the formula, $E_g$ = |Slope| x 1.986 x $10^{-4}$ eV = ……. eV.

**Results**:

| Diode | Cut-in voltage (V) | Static Resistance ($R_s$) (Ω) | Dynamic Resistance ($R_d$) (Ω) | Energy gap (eV) |
|---|---|---|---|---|
| Germanium | | | | |

**Conclusions**:

The forward and reverse bias V-I characteristics of the Germanium diode were studied. The cut-in voltage of the Germanium diode was found from the forward bias characteristics. The reverse saturation current ($I_s$) was found and it was observed that the $I_s$ does not depend on the applied reverse voltage. The energy gap of the semiconductor can be found by operating the PN Junction Diode in the reverse bias mode, and measuring the reverse saturation current as a function of the temperature. The negative slope of the $Log_{10}$ $I_s$ versus 1/T for the fixed reverse bias voltage suggests that the resistance of the semiconductor increases with decreasing temperature.

**Precautions**:

1) Ensure the zero position of the potentiometer of the voltage source before switching ON the power supply.

2) Ensure the connections, and the appropriate voltage and current ranges in the meters before the start of the measurements as well throughout the experiment.

3) Ensure the reading of the zero voltage in each set of measurements.

4) Note the cut-in voltages and appropriate scales with their units on the graph papers. Occupy more than 50% of the graph paper to ensure good presentation of the results.

5) Care must be taken to handle the heating assembly.

6) The temperature must be recorded closest to the Diode without thermometer touching it.

7) The temperature overshoot must be handled by switching off the heating mantle 5 degree before it reaches to 90 ºC.

8) The suggested readings of the voltages and the temperature in the observation table may be suitably adjusted if required for the experiment. Any 6 temperature readings may be taken with a minimum difference of 2 degrees.

9) Connecting wires should be removed carefully and tied after the finish of the experiment.

**Experiment No 6**

## Input and Output characteristics of the NPN transistor in common base configuration

**Aim**:

1. To study the input and output V-I characteristics of the given NPN transistor in the common base (CB) configuration

2) To calculate the input resistance and the current transfer gain in the CB configuration of the NPN transistor.

**Apparatus**: Given Experimental Kit of the NPN transistor, Regulated power supplies, Ammeters (Range 0 to 50 mA and 0 to 25 mA), Milli-Voltmeter (0 to 1000 mV for the input side) and Voltmeter (0 to 20 Volts attached on the output side dc source), Connecting wires.

**Circuit Diagram**: **Refer to the supplementary material for the V-I characteristics**

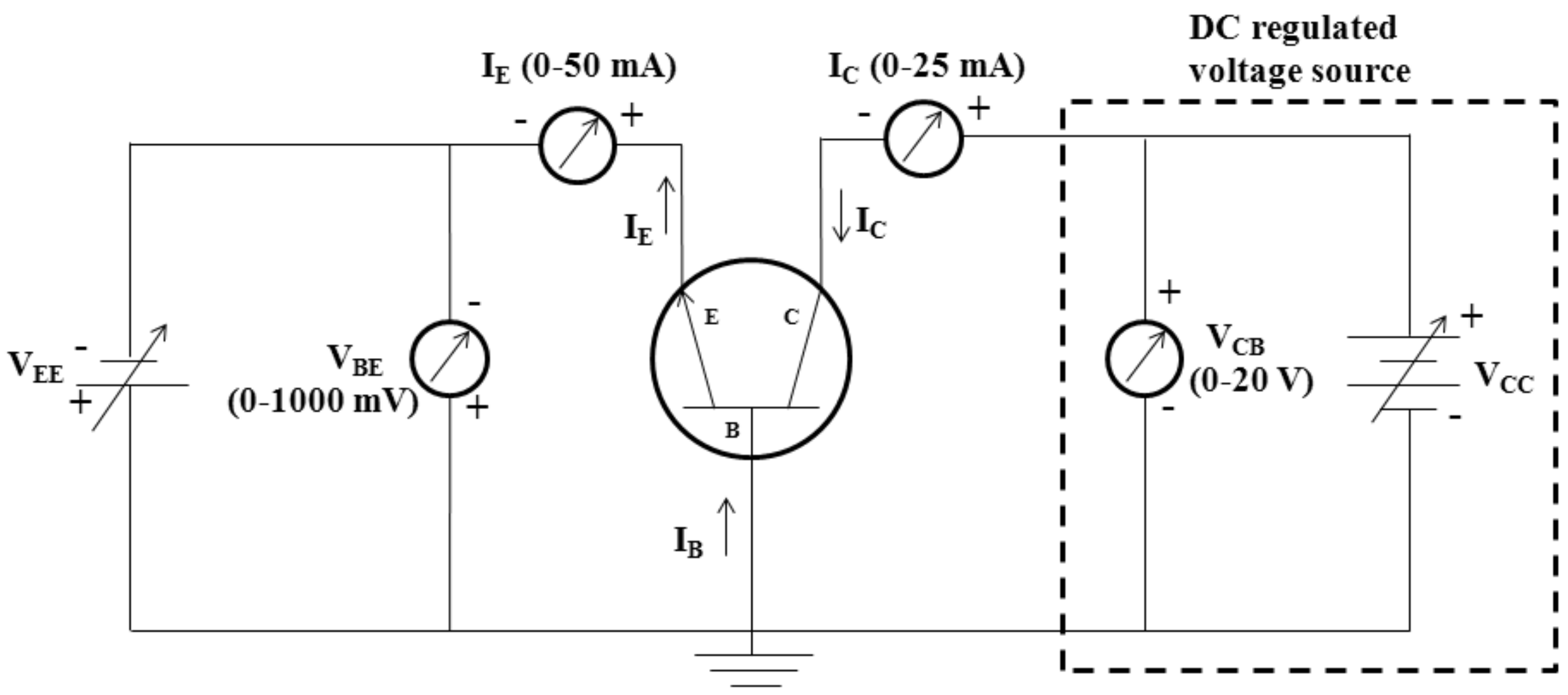

**Formulae:**

1. Transistor parameters in CB mode

(a) $R_i = (V_{BE2} – V_{BE1}) / (I_{E2} – I_{E1})$ at $V_{CB}$ = constant

(b) $\alpha = (I_{C2} – I_{C1}) / (I_{E2} – I_{E1})$ at $V_{CB}$ = constant

where,

$R_i$ = Input resistance,

$V_{BE}$ = Base Emitter voltage,

$V_{CB}$ = Collector Base voltage,

$I_E$ = Emitter current,

$I_C$ = Collector current,

$\alpha$ = Current transfer gain in CB configuration

2. Estimation of error

(a) $y = f(x_1,x_2)$

$\Delta y = |\partial f / \partial x_1| \Delta x_1 + |\partial f / \partial x_2| \Delta x_2$

(b) % error in y = $(\Delta y/y) \times 100$

where,

$x_1$, $x_2$ are the independent variables, and y is the dependent variable.

$\Delta x_1$ and $\Delta x_2$ are determined from the least counts of the instruments

$\Delta y$ is the error in the resulting quantity of an experiment

(c) For input resistance of the NPN transistor in CB mode,

$$R_i = (V_{BE2} – V_{BE1})/(I_{E2} – I_{E1})$$

$$\Delta R_i = 2 \{[1/(I_{E2} – I_{E1})] \Delta V_{BE} + [(V_{BE2} – V_{BE1})/(I_{E2} – I_{E1})^2] \Delta I_E\}$$

where, $\Delta V_{BE}$ = (1/2) L.C. of the millivoltmeter

$\Delta I_E$ = (1/2) L.C. of the milliammeter

**Procedure**:

**1. Input characteristics**

(a) Connect the given NPN transistor in the CB configuration. Connect the milliammeters and the millivoltmeter as shown in the circuit diagram. Use the millivoltmeter for measuring the input voltage ($V_{BE}$) and the milliammeter for measuring the current ($I_E$).

(b) Make $V_{CB}$ = 0 V with the help of the potentiometer on the voltage source ($V_{CC}$).

(c) Increase the base emitter voltage $V_{BE}$ in steps of 0.1 V using the fine knob on the regulated voltage source ($V_{EE}$) and note the corresponding emitter current $I_E$ in mA. Choose the voltage intervals suggested in the table 1.

(d) Increase the collector-base voltage $V_{CB}$ to 6 V and 12 V, and repeat the step (c).

(e) Plot the input V-I characteristics on the graph paper for $V_{CB}$ = 6 V. Choose the origin on the left edge of the graph paper and note the scale (Refer the supplementary material). Determine the cut-in voltage of the Base-Emitter junction and estimate the input resistance $R_i$ from the curve with $V_{CB}$ = 6 V.

**2. Output characteristics**

(a) Increase the input voltage $V_{BE}$ using the fine knob of the voltage source ($V_{EE}$) and make $I_E$ = 4 mA and keep it constant while measuring $V_{CB}$ versus $I_C$.

(b) Increase the $V_{CB}$ in the steps of 1 V using the coarse knob on the voltage source ($V_{CC}$) and note the corresponding collector current $I_C$ in mA. Choose the voltage intervals shown in the table 2. For each reading ensure that the $I_E$ remains constant to the desired value.

(c) Increase $I_E$ to 8 mA, and then to 12 mA, and repeat the step (c).

(d) Plot the output V-I characteristics on the graph paper and write the scale in the top right corner (Refer the supplementary material). Choose the origin on the bottom left corner of the graph paper. Calculate the current transfer gain at $V_{CB}$ = 2 V.

**Observation Tables**:

**Least Counts (L.C.) of the measuring instruments**

(1) Milli-ammeter (for $I_E$) with Range (………..): L.C. = 1 Main scale reading = ……

(2) Milli-voltmeter (for $V_{BE}$) with Range (………): L. C. = 1 Main scale reading = ……

(3) Milli-ammeter (for $I_C$) with Range (………..): L.C. = 1 Main scale reading = ……

(4) Voltmeter (for $V_{CB}$) with Range (………): L. C. = 1 Main scale reading = ……

**1. Input V-I characteristics**

| S. No. | $V_{CB}$ = 0 V | | $V_{CB}$ = 6 V | | $V_{CB}$ = 12 V | |
|---|---|---|---|---|---|---|
| | $V_{BE}$ (V) | $I_E$ (mA) | $V_{BE}$ (V) | $I_E$ (mA) | $V_{BE}$ (V) | $I_E$ (mA) |
| 1 | 0 | | 0 | | 0 | |
| 2 | 0.1 | | 0.1 | | 0.1 | |
| 3 | 0.2 | | 0.2 | | 0.2 | |
| 4 | 0.3 | | 0.3 | | 0.3 | |
| 5 | 0.4 | | 0.4 | | 0.4 | |
| 6 | 0.5 | | 0.5 | | 0.5 | |
| 7 | 0.55 | | 0.55 | | 0.55 | |
| 8 | 0.6 | | 0.6 | | 0.6 | |
| 9 | 0.65 | | 0.65 | | 0.65 | |
| 10 | 0.7 | | 0.7 | | 0.7 | |

**2. Output V-I Characteristics**

| S. No. | $I_E$ = 4 mA | | $I_E$ = 8 mA | | $I_E$ = 12 mA | |
|---|---|---|---|---|---|---|
| | $V_{CB}$ (V) | $I_C$ (mA) | $V_{CB}$ (V) | $I_C$ (mA) | $V_{CB}$ (V) | $I_C$ (mA) |
| 1 | 0 | | 0 | | 0 | |
| 2 | 1 | | 1 | | 1 | |
| 3 | 2 | | 2 | | 2 | |
| 4 | 3 | | 3 | | 3 | |
| 5 | 4 | | 4 | | 4 | |

**Calculations:**

1. From input characteristics, Slope = $(I_{E2} - I_{E1})/ (V_{BE2} - V_{BE1})$ = …..

Input resistance, $R_i$ = 1/slope = …….Ω

2. Current gain, $\alpha = (I_{C2}-I_{C1})/ (I_{E2}-I_{E1}) \approx$ ….. at $V_{CB}$ = 2 volts

3. Error in $R_i$,

(a) $\Delta R_i = 2 \{[1/(I_{E2} - I_{E1})] \Delta V_{BE} + [(V_{BE2} - V_{BE1})/(I_{E2} - I_{E1})^2] \Delta I_E\}$

where, $\Delta V_{BE}$ = (1/2) L.C. of the voltmeter

$\Delta I_E$ = (1/2) L.C. of the ammeter

(b) % error in y = $(\Delta R_i/R_i)$ x 100

**Results**:

From the input characteristics, i.e., from the graph of $V_{BE}$ versus $I_E$ at $V_{CB}$ = 0 V, the value of $R_i$ of the given NPN transistor was found to be …. **Ω**. The current gain of the NPN transistor in the CB configuration was found to be, $\alpha \approx$ ……at $V_{CB}$ = 2 volts. The error was estimated in the calculation of the input resistance, and it was around …..%.

**Conclusions**:

The input V-I characteristics of the given NPN transistor were studied. The current gain in the common base configuration was found be close to unity. The current transfers from the lower resistance input side to the higher resistance output side using the transistor action which is not feasible by electrical techniques. Since the current transfer gain is unity, effectively the resistance is transferred from the input side to the output side of the transistor.

**Precautions**:

1) Ensure the zero position of the potentiometer on the voltage sources before switching ON the power supply to the kit.

2) Ensure the connections, and the appropriate voltage and current ranges in the meters before the start of the measurements.

3) Ensure the reading of the zero voltage in each set of measurements.

4) Note the cut-in voltages and appropriate scales with their units on the graph papers. Occupy more than 50% of the graph paper to ensure good presentation of the results.

5) The suggested voltage readings in the observation table may vary slightly, so choose the voltage intervals appropriate for the experimental set-up.

6) Connecting wires should be removed carefully and tied after the finish of the experiment.

**Experiment No 7**

**Determination of the radius of curvature of a plano convex lens using Newton's rings method**

**Aim**: To determine the radius of curvature of a plano convex lens using Newton's rings

**Apparatus**: Sodium vapor lamp, plano convex lens, glass plate, travelling microscope, reading lamp

**Figure: Refer the supplementary material**

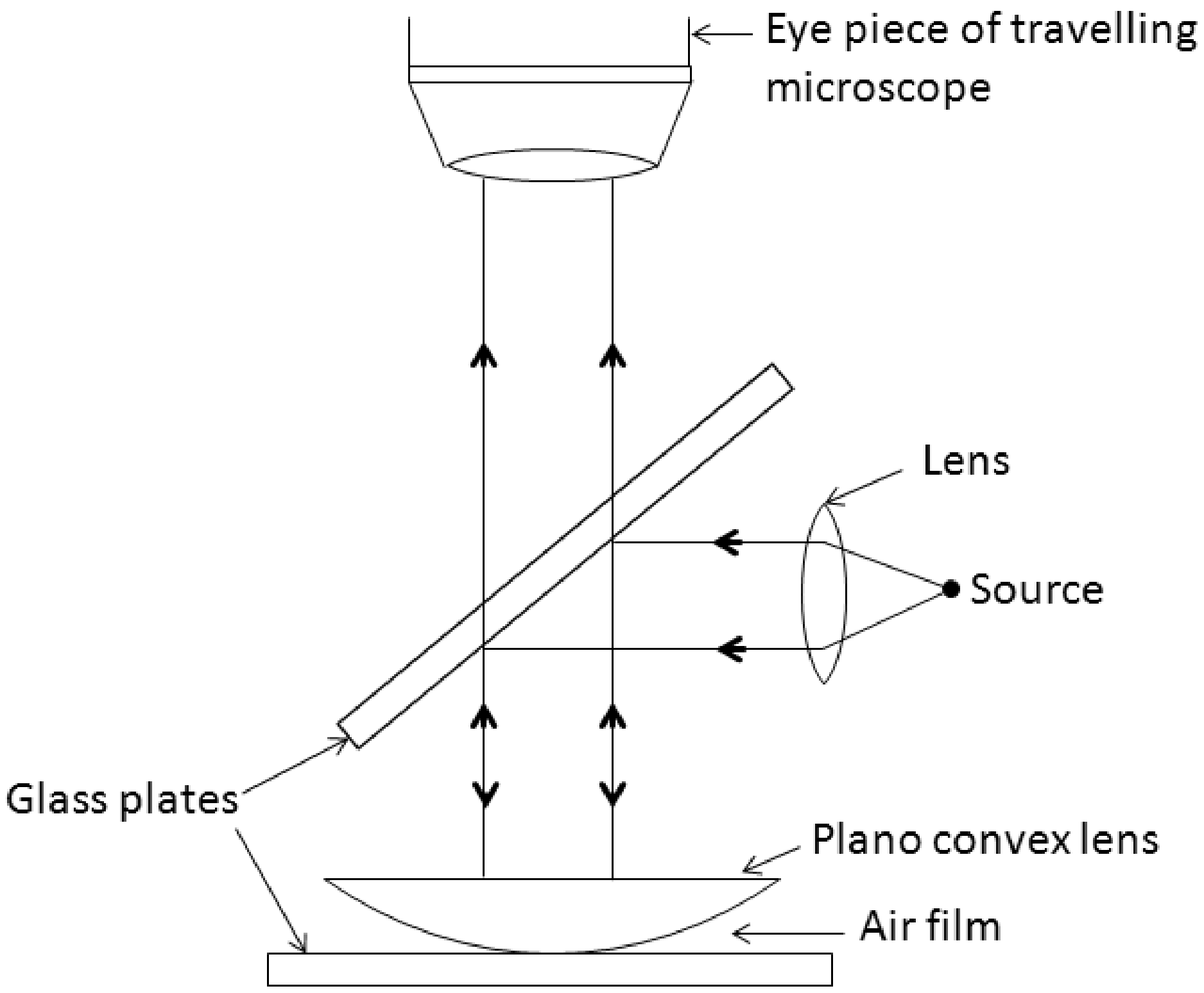

**Formula**

$$R = \text{Slope}/(4\lambda) \qquad \text{cm}$$

where, Slope = $(D_{n+p})^2 - (D_n)^2/[(n+p)-(n)]$

R = radius of curvature of a plano convex lens

$D_{n+p}$ = diameter of $(n+p)^{th}$ ring

$D_n$ = diameter of the $n^{th}$ ring

$\lambda$ = wavelength of monochromatic sodium light

p = Number of rings or the denominator in the calculation of the slope

**Procedure**:

(a) Find the least count of the reading scale on the travelling microscope. Note down the main scale reading of one division and the total number of divisions on the circular scale. Arrange the apparatus and obtain a stable pattern of concentric circular rings. Use the inclined glass plate to get the interference pattern with a dark central spot.

(b) Starting from the first dark ring (minimum diameter), count the $18^{th}$ dark ring as the crosswire moves always parallel to the tangent of the first ring. Note the main scale reading and the coinciding circular scale division.

(c) Gradually move the crosswire to $14^{th}$, $10^{th}$, $6^{th}$, $2^{nd}$ ring and tabulate the MSR and CSD in each case. Move the crosswire on the right side to the first dark ring, and successively note the MSR and CSD for the $2^{nd}$, $6^{th}$, $10^{th}$, $14^{th}$, $18^{th}$ ring. Calculate the total reading of each dark ring position using the formula MSR + CSD x LC.

(d) Subtract the right side readings from the left side readings to get the absolute difference, which gives the diameter of the dark rings. Plot a graph between $(D_n)^2$ versus n, and obtain the slope. Calculate the radius of the plano convex lens using the formula in centimeters.

**Observation Table**:

1. **Least count of the reading scale**

   Smallest main scale reading = …. mm

   Total number of divisions on the circular scale = ….. div

   Least Count = 1 MSR/Total CSD = ……mm

**2. Measurement of the fringe diameters**

| No. of dark fringe n | Readings on L.H.S. | | | Readings on R.H.S. | | | $D_n$ (mm) | $(D_n)^2$ $(mm)^2$ |
|---|---|---|---|---|---|---|---|---|
| | MSR | VSD | TR (mm) | MSR | VSD | TR (mm) | | |
| 2 | | | | | | | | |
| 6 | | | | | | | | |
| 10 | | | | | | | | |
| 14 | | | | | | | | |
| 18 | | | | | | | | |

**Calculations**: Given, wavelength of the sodium light, $\lambda$ = 5890 Å

1) Slope = $[(D_{n+p})^2 - (D_n)^2] / [(n+p) - (n)]$ = …..

2) Radius of the plano convex lens, R = Slope / 4$\lambda$ = …….. cm

**Result**: Square of the fringe diameter varies linearly with the number of the dark fringe.

The radius of curvature of the given plano convex lens was found to be ….cm.

**Conclusions**: The Newton's rings were obtained using a plano convex lens placed on a glass plate, effectively producing a wedge shaped film with the varying thickness. The radius of the plano convex lens was found by measuring experimentally the fringe diameters, and it was found to be…….cm in the given experiment.

**Precautions**:

1) The crosswire must be tangential to the dark fringes, and must move along the diameter.

2) The readings should be noted only in one direction of movement of the crosswire in order to avoid the backlash error. If more than 3 readings are missed, then the measurements should be repeated entirely.

3) The numbers of dark rings in the observation table 2 are suggested readings. In the experiment any 5 rings can be used for measuring the diameter, with a minimum difference of 2.

**Experiment No 8**

## Determination of the wavelength of sodium light using plane diffraction grating

**Aim**: To determine the wavelength of sodium light using plane transmission grating

**Apparatus**: Spectrometer, Grating, Sodium vapor lamp, Reading lens

**Figure: Refer to the supplementary material**

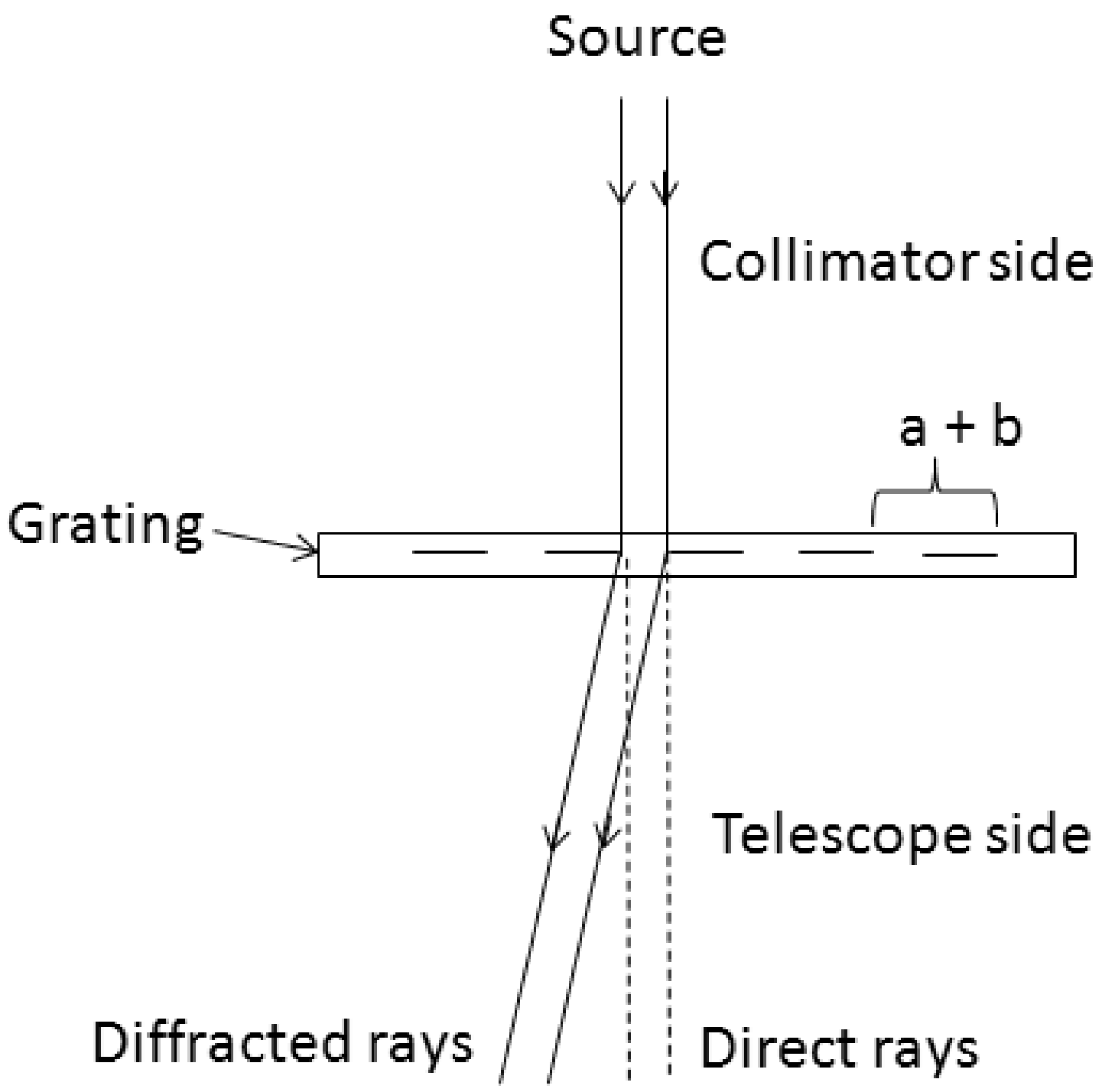

**Formula**

1. Wavelength of the sodium light, $\lambda = d\sin(\Theta_n)/n$ Ǻ

where, $\Theta_n$ = Angle of diffraction for $n^{th}$ order

n = order of spectrum

d = grating element (a+b), i.e. the width of one transparent region and one opaque region

2. The error in the estimation of the wavelength, $\Delta\lambda = 2(d/n)(\cos\theta_n)\Delta\theta$ Ǻ

where, $\Delta\theta$ = Least Count of the spectrometer

3. Percentage error, $\% \lambda = 100 \times \Delta\lambda/\lambda$

**Procedure**:

**1. Study of the spectrometer**

(a) Note different parts of the spectrometer, e.g., collimator, slit (open and close), telescope, grating table, height adjusting screws, grating table plate, different knobs to fix the positions of the telescope and the grating table plate and the fine movement knob of the telescope. Note that the movement of the telescope changes the readings in the glass windows (W1 as well as W2) due to the circular scale (also called main scale). Moving the grating table plate changes the readings due to the movement of the vernier scale. Refer Appendix to note the construction of the spectrometer as well as the reading scales.

(b) Find the least count of the reading scale on the spectrometer. Note down the main scale reading of one division and the total number of divisions on the vernier scale. Calculate the total reading of for each vertical cross-wire position (in the eye-piece of the telescope) using the formula MSR + VSD x LC. Any window can be labeled as W1 and the other as W2. Diametrically opposite placement of the windows results in the difference of 180 degrees for the total reading of the vertical crosswire position.

**2. Setting up of the experiment**

(a) Adjust the width of the slit so that the light from the sodium source enters the collimator.

(b) Bring the telescope in line with the collimator and focus the collimator as well as telescope on the slit to obtain a thin and sharp image. Use the slit width control knob, collimator focus knob and the telescope focus knob.

(c) Position the grating on the table so that the free surface of the grating faces the collimator and the windows are positioned nearly parallel to the grating surface. Use the movement of the grating table plate and the grating table for this adjustment.

**3. Setting up the grating for the normal incidence of incoming rays**

(a) Note the reading of the telescope in the direct image position. This corresponds to the angular separation of 180 degrees between the collimator and the telescope and can be used as a reference. Tabulate the MSR from either window W1 or W2 (whichever is opposite to the telescope position) in the journal and calculate the difference of 90 degree.

(b) Rotate the telescope by 90 degree to the right (or to the left) side. Note the MSR from the same window when the telescope is at the right angle (90 degree) position with respect to the collimator.

(c) Rotate slowly the grating table plate (not moving the telescope plate) till you see the reflected image of the slit from the grating surface in the telescope. This position of the grating ensures 45 degree angle between the grating surface and the incoming rays. If the single reflected image is not seen, then adjust the collimator or the telescope focusing knobs if required.

(d) Rotate the grating table plate by 45 degrees on the left (or right), using the scale in either window W1 or W2. This will align the grating surface in line with the telescope and will make the grating surface perpendicular to the incoming rays. Position of the the grating table plate (and effectively the position of the vernier scale) should remain the same during the experiment.

(e) Bring the telescope in line with the collimator by adding 90 degree to the left (or right) and obtain the position of direct slit image on the vertical cross-wire in the eye-piece of the telescope.

**4. Measurement of angle of deviation for the spectral lines D1 and D2 of the sodium source**

(a) Note the direct image position of the slit in the telescope. The readings of MSR and VSD should be noted from both the windows, W1 and W2. Tabulate the readings in the observation table.

(b) Rotate the telescope to the left (or the right) to find the 1$^{st}$ order images of the two spectral lines D1 and D2. Care should be taken so that the telescope rotates without moving the grating table plate.

(c) The diffracted image of the slit, at successive positions, is labeled as D1 and D2. Use the telescope and collimator focusing knobs to obtain fine diffracted images of the slit.

(d) Continue to rotate the telescope to find the 2$^{nd}$ order diffraction spectra. Use the focus knob to clearly observe the two fine images of the slit.

(e) Keep the vertical cross-wire on the image D1 first and note the position using the MSR and VSD of the spectrometer windows, W1 and W2.

(f) Use the fine knob below the telescope to move the cross-wire from D1 to D2 and note the readings of MSR and VSD from the windows.

(g) Calculate the angle of deviation for both the spectral lines with respect to the direction image readings. Follow the observation table to find the differences from both the windows. Calculate the wavelengths of sodium spectral lines, D1 and D2, using the formula.

**Observation Table**:

**1) Least count (L.C.) of the reading scale**

Smallest main scale reading = ……

Total number of divisions on the circular scale = …..

L.C. = 1 MSR/Total VSD = ……

**2) Angle of deviation of D1 and D2**

| Spectral line | Deviated spectral line | | | | | | Direct image of the slit | | | | | | $\Theta$ | Mean $\Theta$ |
|---|---|---|---|---|---|---|---|---|---|---|---|---|---|---|
| | W1 | | | W2 | | | W1 | | | W2 | | | | |
| | MSR | VSD | TR | MSR | VSD | TR | MSR | VSD | TR | MSR | VSD | TR | | |
| D1 | | | | | | | | | | | | | $(\Theta)_{W1}$<br>$(\Theta)_{W2}$ | |
| D2 | | | | | | | | | | | | | $(\Theta)_{W1}$<br>$(\Theta)_{W2}$ | |

**Calculations**: Given, $d = a + b = 2.54$ cm/15000 = …….cm

1) For spectral line D1, n=2, $\lambda = d\sin(\Theta_n)/n$ = …… Å

2) For spectral line D2, n=2, $\lambda = d\sin(\Theta_n)/n$ =……. Å

**Result**:

| Parameters | Angle of diffraction | Spectral wavelength |
|---|---|---|
| For spectral line D1 | | |
| For spectral line D2 | | |

**Conclusions**: The Diffraction grating was used to obtain the spectral lines of the sodium light source, D1 and D2 as…...nm and…..nm, respectively.

**Precautions**:

1) The crosswire must be on the center position of the spectral line.

2) The grating table should be leveled so that the slit image is at the center position of the view of the telescope.

3) The spectral lines should be fine and bright. Backlash error should be avoided.

**Self-study experiment**

**Method of Linear least square fitting**

**Aim**: To determine the parameters of best fit and plot of the line of best fit for the data of the energy gap experiment

**Usables**: Experimental data of the experiment of energy gap of semiconductor, Calculator, Graph paper

**Formula**:

**1. Least Square Fitting**

$$m = [N \sum_i x_i y_i - \sum_i x_i \sum_i y_i ] / [N \sum_i x_i^2 - (\sum_i x_i)^2] \quad (1)$$

$$c = [\sum_i x_i^2 \sum_i y_i - \sum_i x_i \sum_i x_i y_i ] / [N \sum_i x_i^2 - (\sum_i x_i)^2] \quad (2)$$

$$y_{ls} = m x_i + c \quad (3)$$

$$E = \sum_i (y_i - y_{ls})^2 \quad (4)$$

Where, $m$ = slope of the line of best fit,

$c$ = intercept of the line of best fit,

$y_i$= the observed value of y at $x = x_i$,

$y_{ls}$= the corresponding value on the line of the best fit and

N = the total number of observations.

**2. Energy gap**,

Energy gap, $E_g = |m| \times 1.986 \times 10^{-4}$eV

**Procedure**:

(a) Use N = 11, which is the number of experimental observations in the experiment of energy gap of semiconductor.

(b) Using the experimental data, calculate the values of $\sum_i x_i$,$\sum_i y_i$, $\sum_i x_i^2$ and $\sum_i x_i y_i$. Use the observation table I.

(c) Calculate the values of m and c using the formulae 1 stated above.

(d) Calculate the values of $y_{ls}$ using the corresponding values of $x_i$.

(e) Calculate the differences ($y_i$-$y_{ls}$), which are differences between the experimental values $y_i$ and the corresponding best fit values $y_{ls}$ for different values of $x_i$.

(f) Calculate the squares of errors, i.e., $E_i = (y_i-y_{ls})^2$ and their sum $\sum_i E_i^2$.

(g) Change the value of m by 1% and repeat the steps (d) to (f). Use the value of c in step (c).

(h) Change the value of c by 1% and repeat the steps (d) to (f). Use the value of m in step (c).

(i) Plot the best fit line using the $x_i$ and $y_{ls}$ calculated in the step (d).

(j) Use the value of m in the step (c) to calculate the energy gap of the semiconducting material of the PN Junction diode.

**Observation Table**:

1. Calculation of the best fit parameters and least square error

| S. No | $x_i$ = 1/T ($K^{-1}$) | $y_i = Log_{10} I_s$ | $(x_i)^2$ | $x_iy_i$ | $y_{ls} = mx_i+c$ | $E_i = y_i-y_{ls}$ | $E_i^2 = (y_i-y_{ls})^2$ |
|---|---|---|---|---|---|---|---|
| 1 | 0.003 | 1.69897 | | | | | |
| 2 | 0.00302 | 1.66276 | | | | | |
| 3 | 0.00304 | 1.62325 | | | | | |
| 4 | 0.00306 | 1.5563 | | | | | |
| 5 | 0.00308 | 1.49136 | | | | | |
| 6 | 0.0031 | 1.44716 | | | | | |
| 7 | 0.00312 | 1.36173 | | | | | |
| 8 | 0.00314 | 1.30103 | | | | | |
| 9 | 0.00317 | 1.23045 | | | | | |
| 10 | 0.00321 | 1.07918 | | | | | |
| 11 | 0.00324 | 1 | | | | | |
| Sums | $\sum_i x_i$ = …. | $\sum_i y_i$ = ….. | $\sum_i x_i^2$ = ….. | $\sum_i x_iy_i$ = ……. | | | $\sum_i E_i^2$ = ….. |

2. Verification of least square error by changing (a) m by 1% and then (b) c by 1%

| S. No | (a) $y_{ls} = mx_i+c$ | (a) $E_i = y_i-y_{ls}$ | (a) $E_i^2 = (y_i-y_{ls})^2$ | (b) $y_{ls} = mx_i+c$ | (b) $E_i = y_i-y_{ls}$ | (b) $E_i^2 = (y_i-y_{ls})^2$ |
|---|---|---|---|---|---|---|
| 1 | | | | | | |
| 2 | | | | | | |
| 3 | | | | | | |
| 4 | | | | | | |
| 5 | | | | | | |
| 6 | | | | | | |
| 7 | | | | | | |
| 8 | | | | | | |
| 9 | | | | | | |
| 10 | | | | | | |
| 11 | | | | | | |
| Sums | | | $\sum_i E_i^2$ = ….. | | | $\sum_i E_i^2$ = ….. |

**Calculations:**

1. Best fit parameters,

$m = [N \sum_i x_i y_i - \sum_i x_i \sum_i y_i ] / [N \sum_i x_i^2 - (\sum_i x_i)^2]$ = ……..

$c = [\sum_i x_i^2 \sum_i y_i - \sum_i x_i \sum_i x_i y_i ] / [N \sum_i x_i^2 - (\sum_i x_i)^2]$ = …….

2. Energy gap,

$E_g = |m| \times 1.986 \times 10^{-4}$ eV

= …………………..eV

**Results**: The parameters of the best fit were found to be m =…..and c =……. The line of best fit was plotted using these parameters. The energy gap of the semiconductor material of the PN Junction diode was calculated using the slope.

**Conclusions**:

The least square method was studied to find the best fit line which describes the experimental data points on the graph. The slope and the intercept of the best fit line were found from the least square method formulae. The energy gap of the semiconducting material of the given PN Junction diode was calculated using the slope of the best fit line. The verification of the minimum error was done by varying arbitrarily the slope and the intercept value by 1 %, and it was ascertained that the sum of the squares of errors increases for parameters other than found by the least square formulae.

**Precautions**:

1) Care must be taken to correctly calculate the sums for the least square formulae.
2) The sums and the values of the parameters should be determined only upto 4 significant digits, and this has to be consistently maintained throughout the calculations.
3) Note that the method of the least square is a statistical procedure, and it does not give any meaningful information on the actual value of the result obtained via experiment.

**Appendix**:

**1. V-I Characteristics of the photocell**

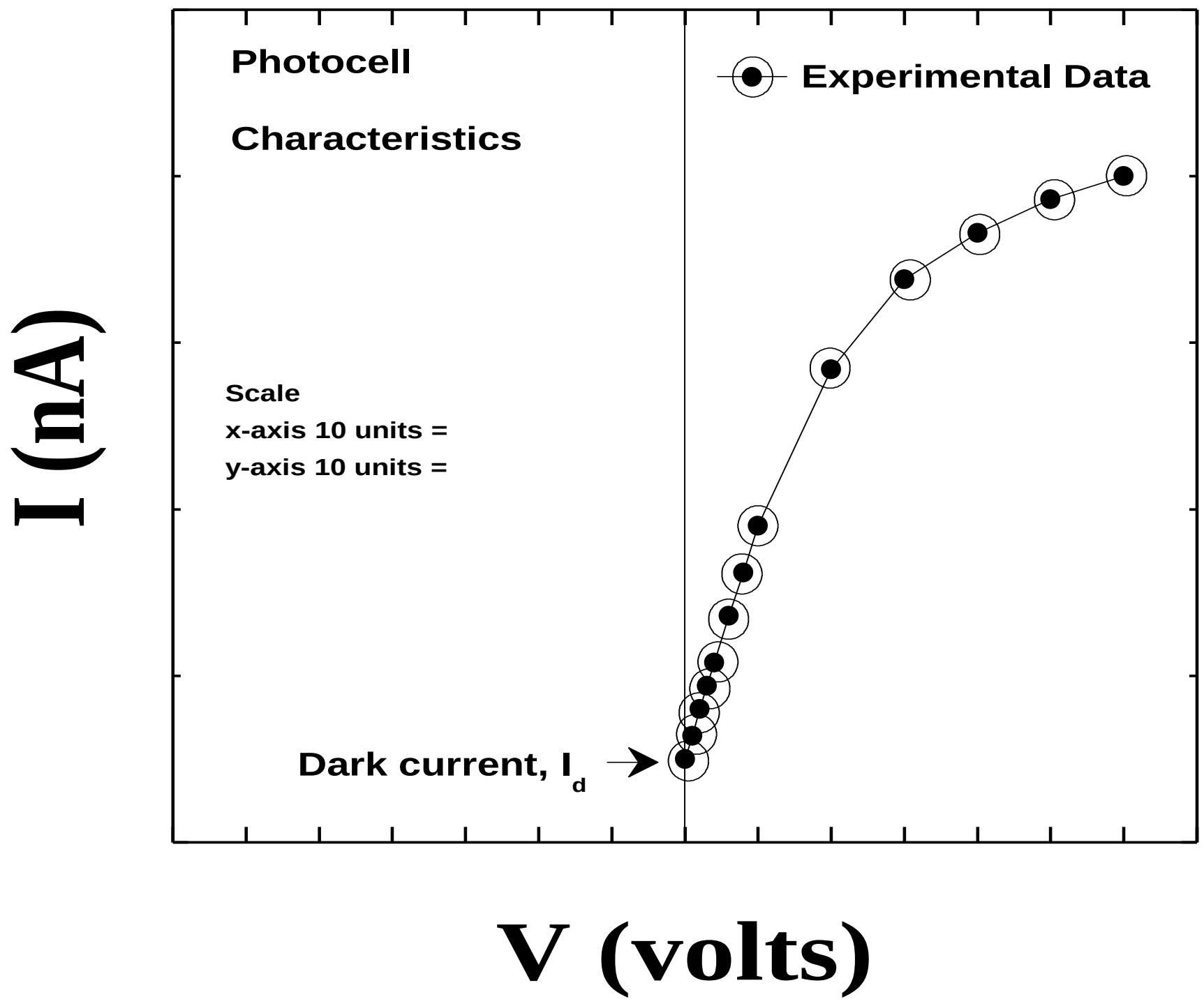

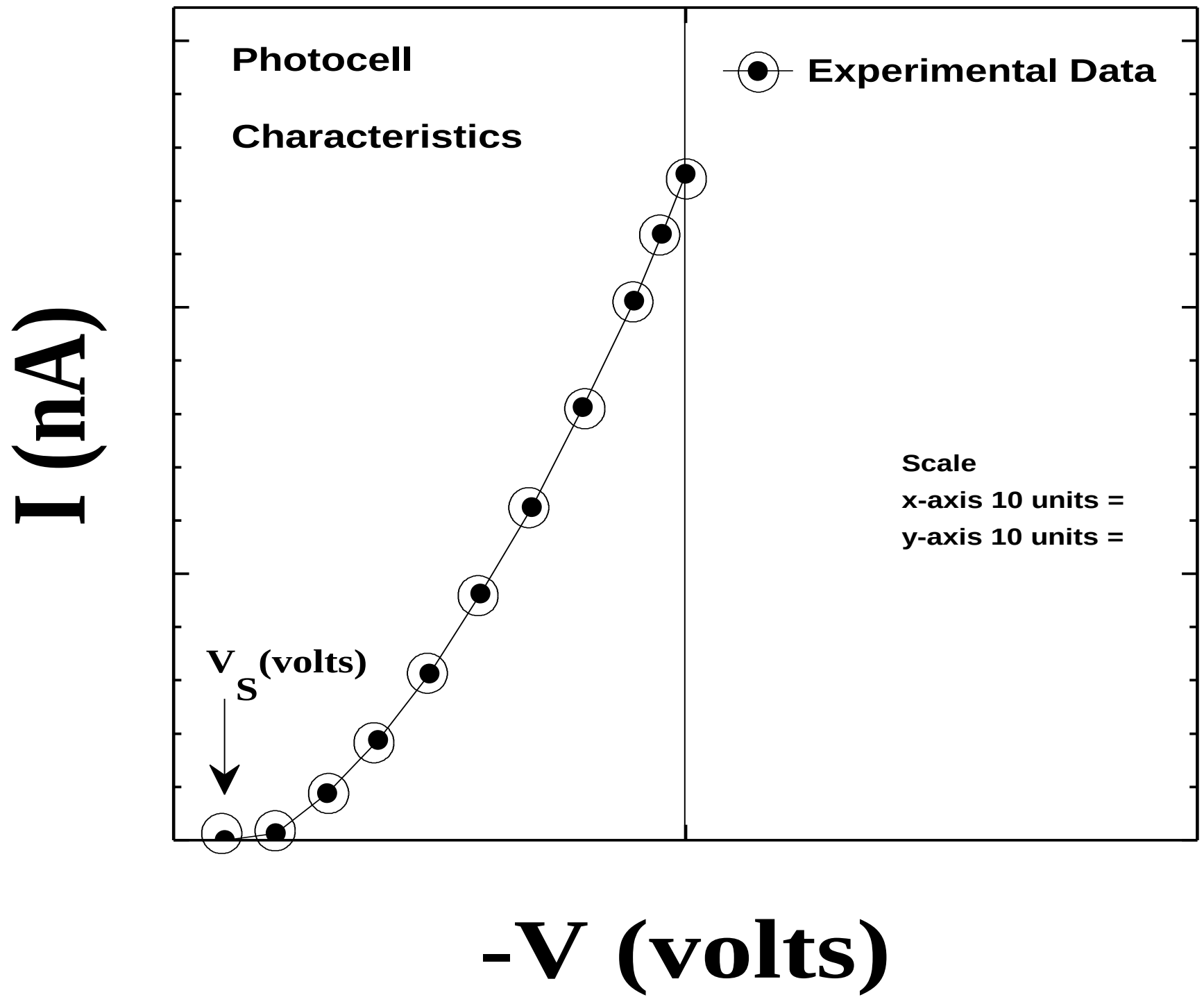

**2. Study of Hall Effect**

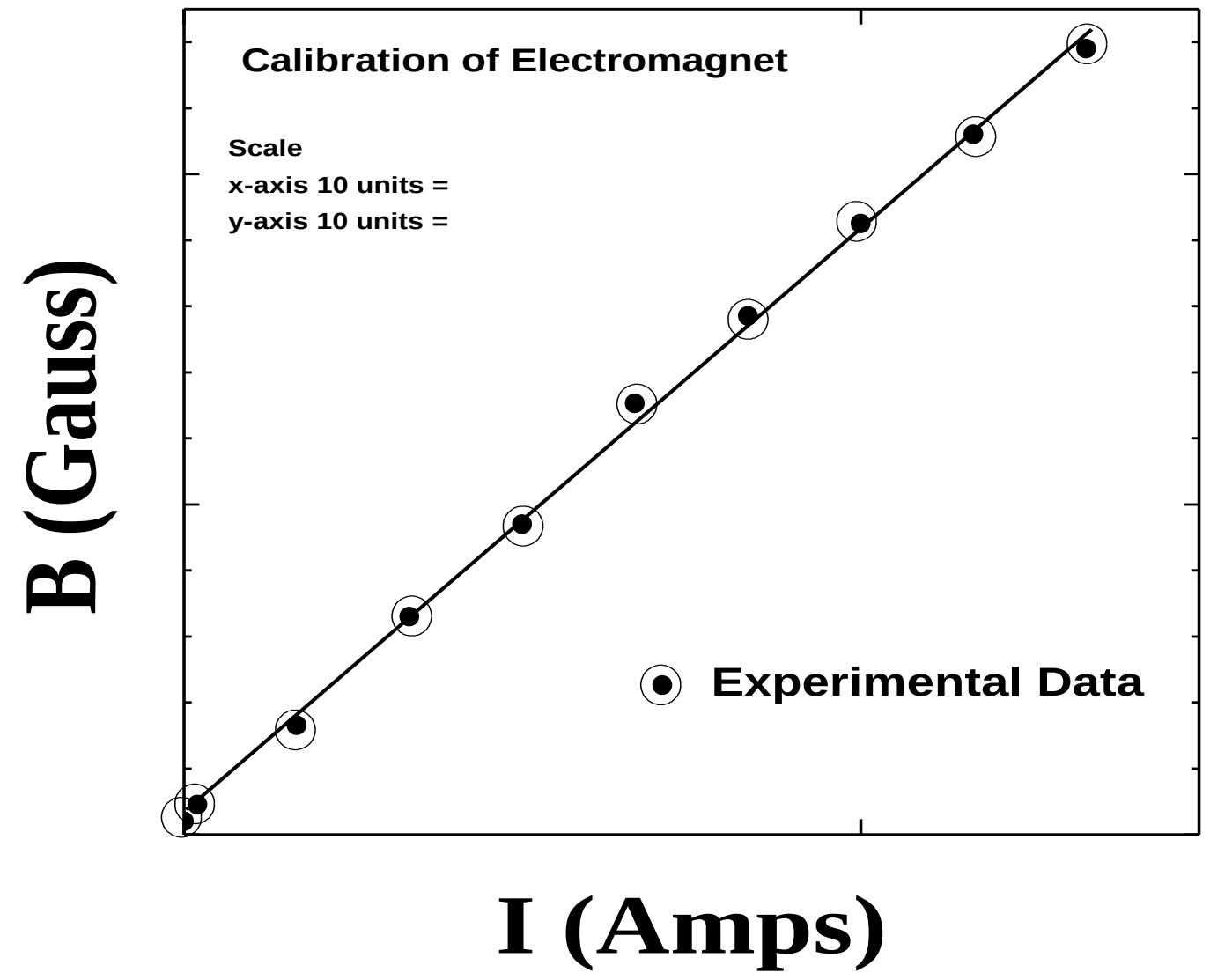

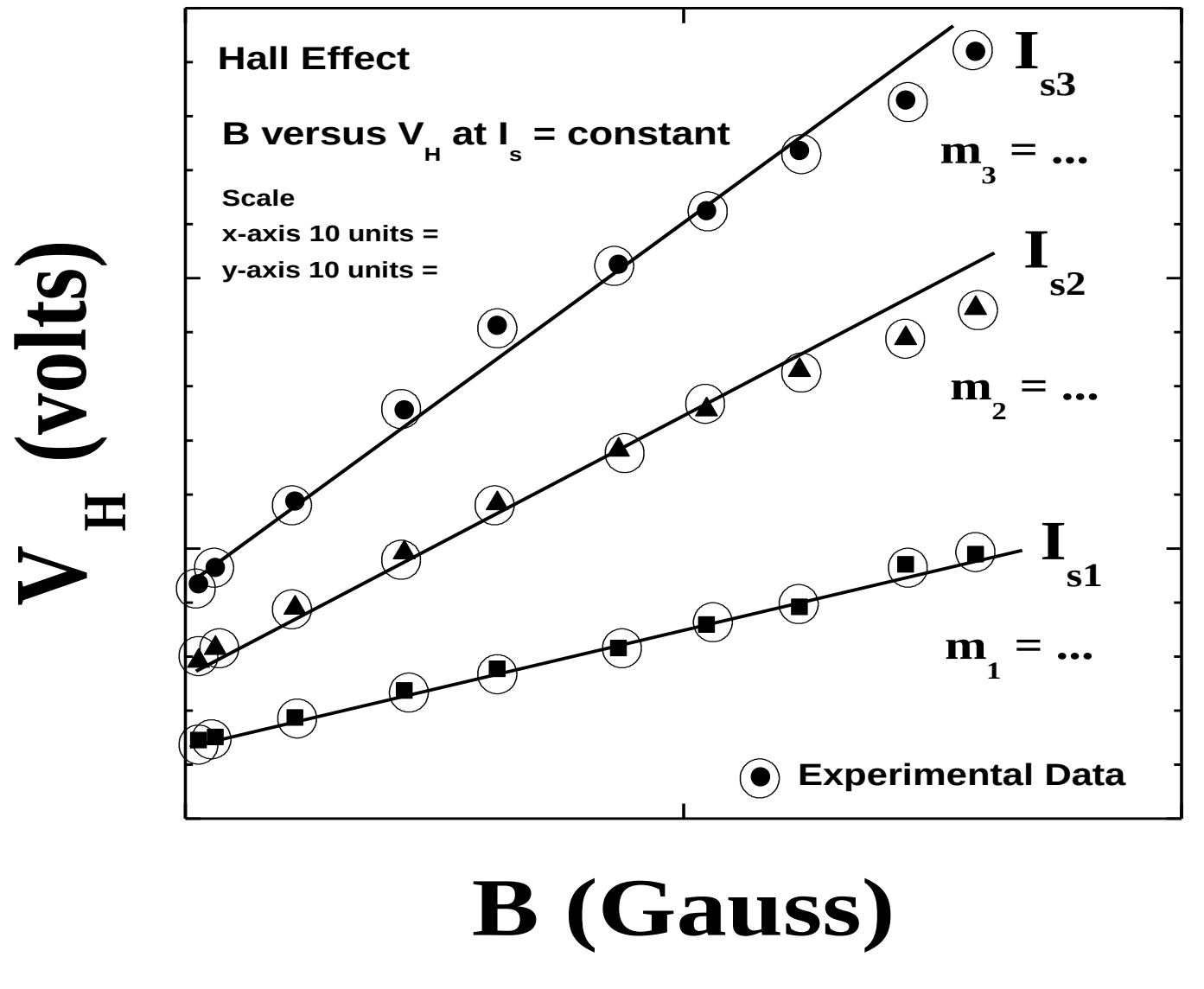

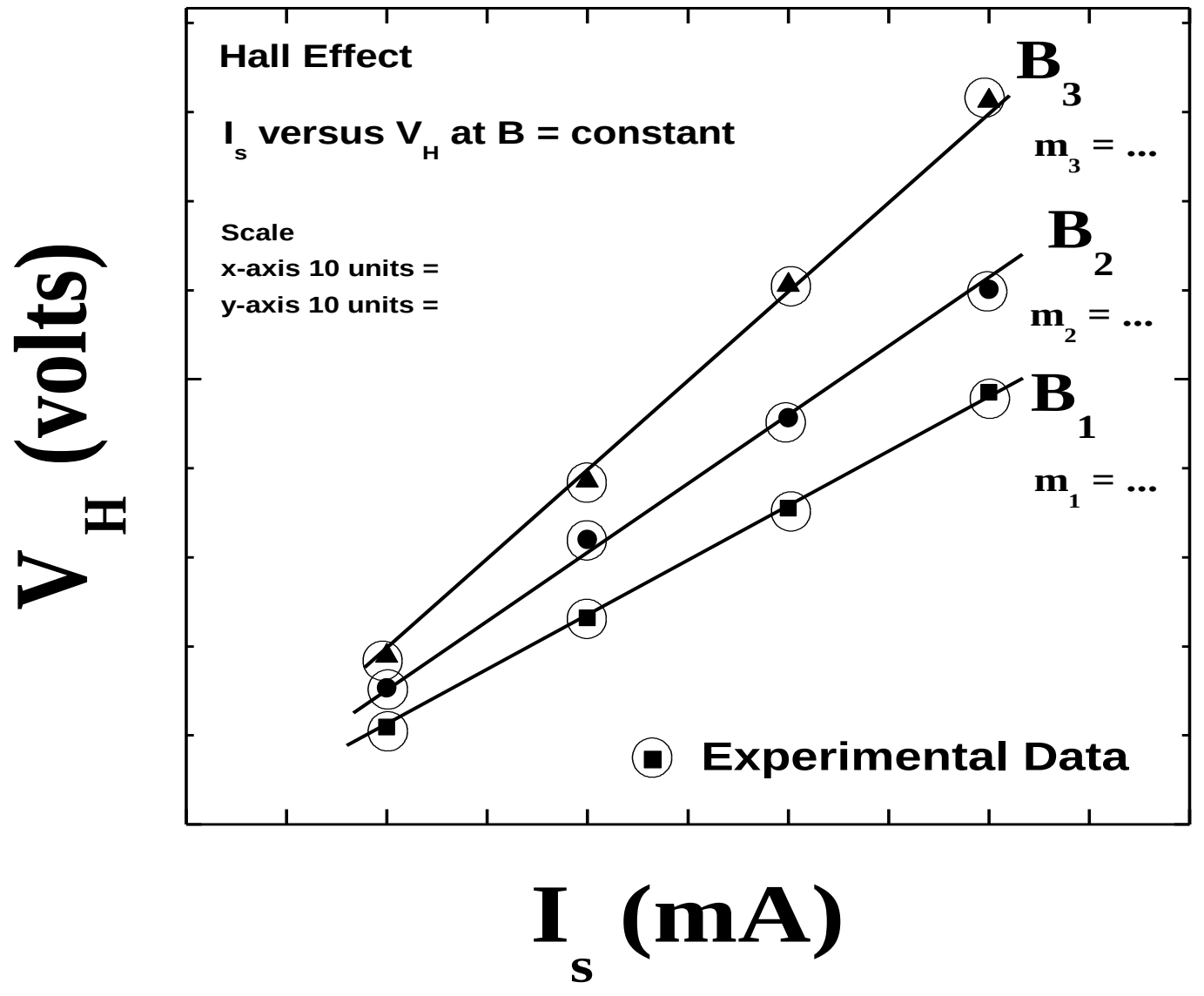

**3. Determination of e/m by Thomson's method**

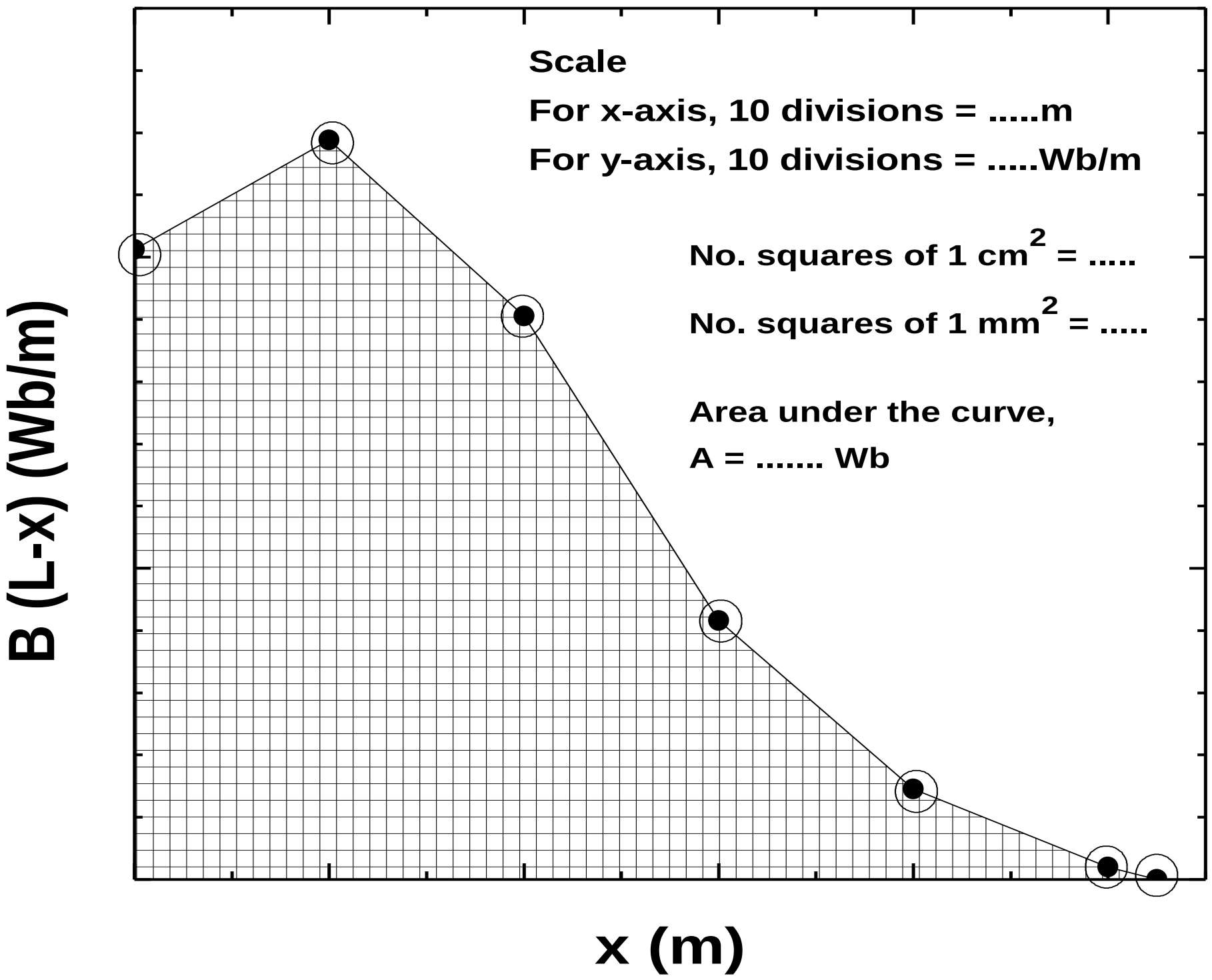

# 4. Study of Cathode Ray Oscilloscope

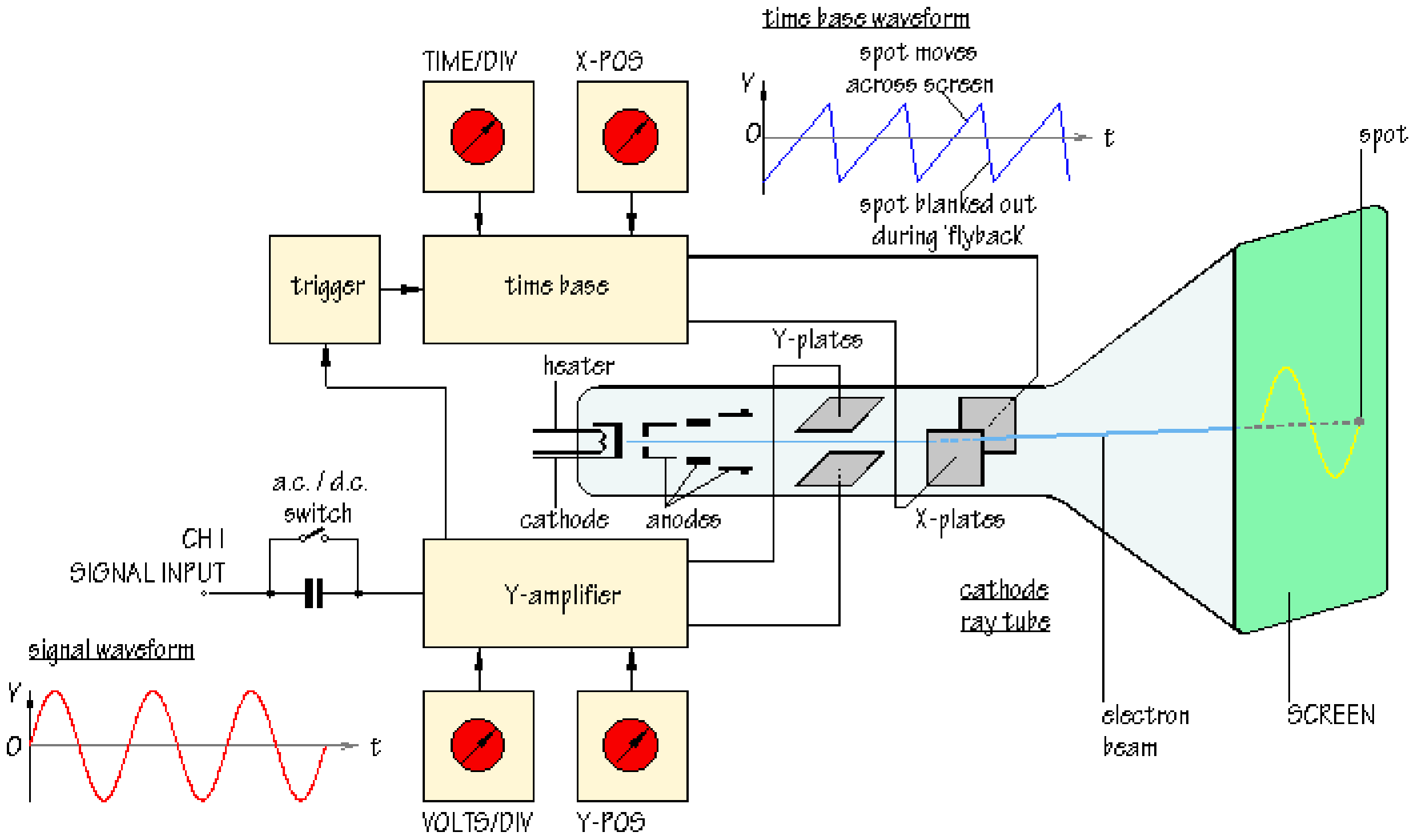

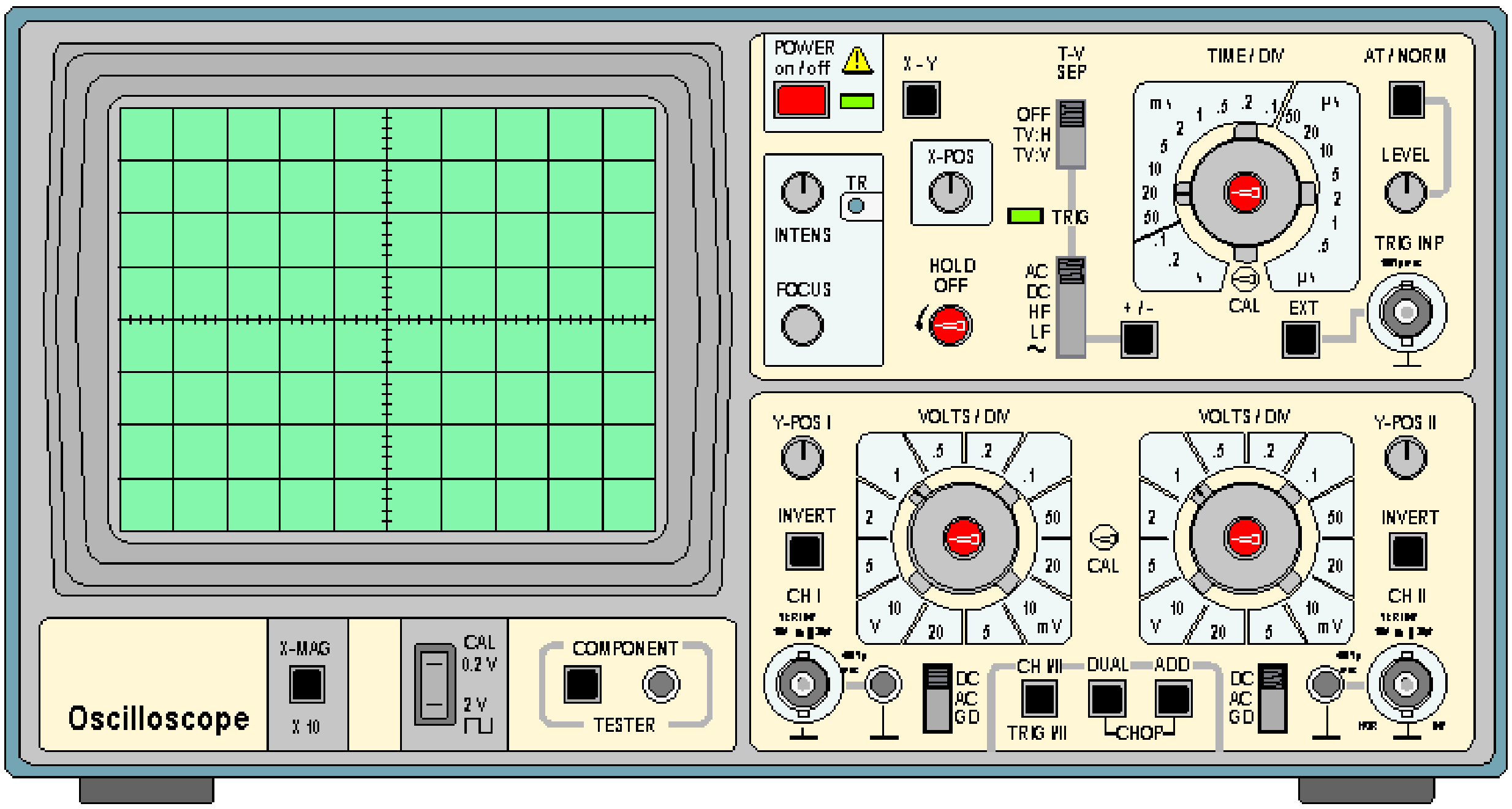

**4a. Phase determination**

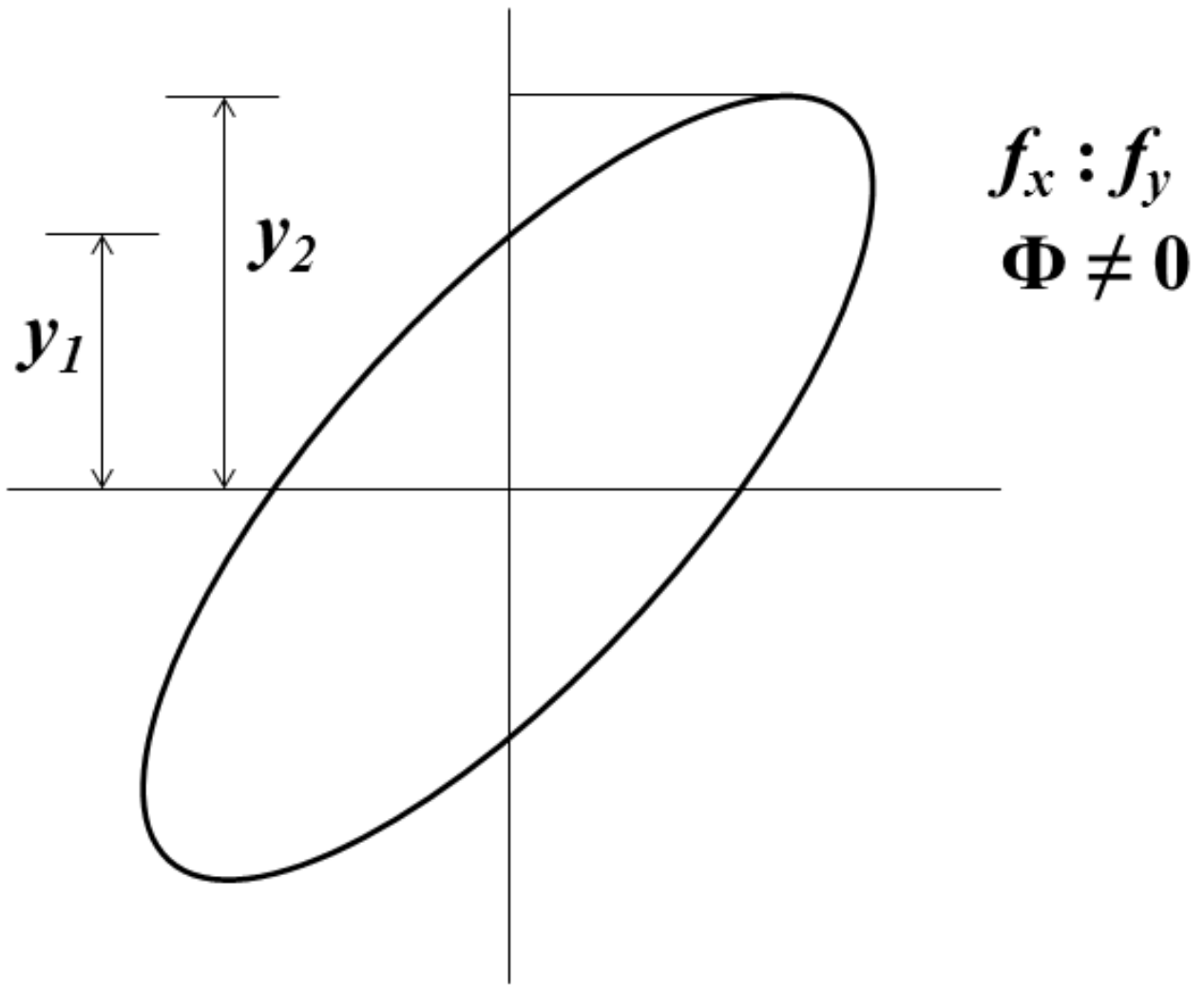

**4b. Lissajous patterns**

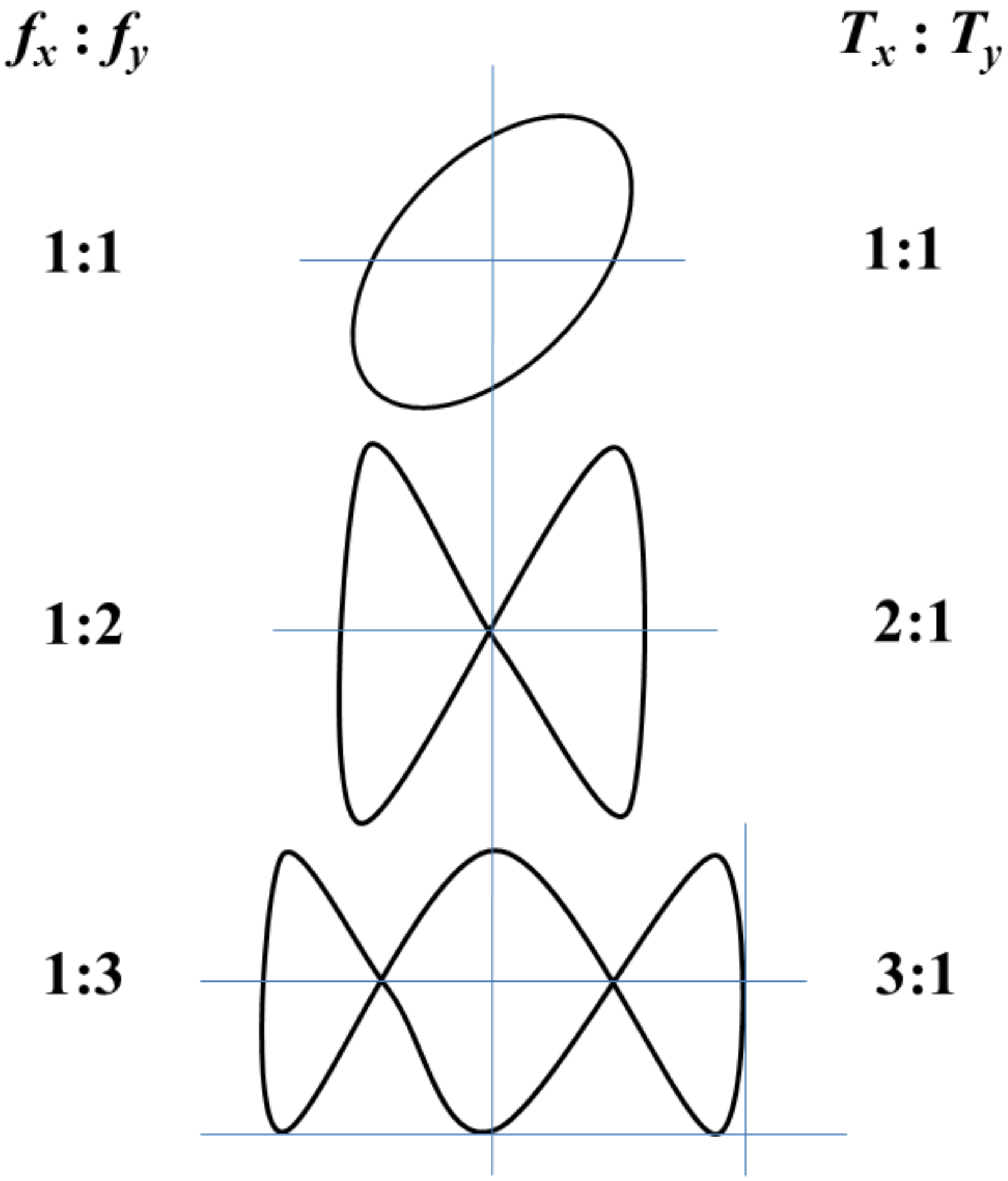

**5. Study of semiconductor diode**

**a. V-I Characteristics of the Semiconductor diodes**

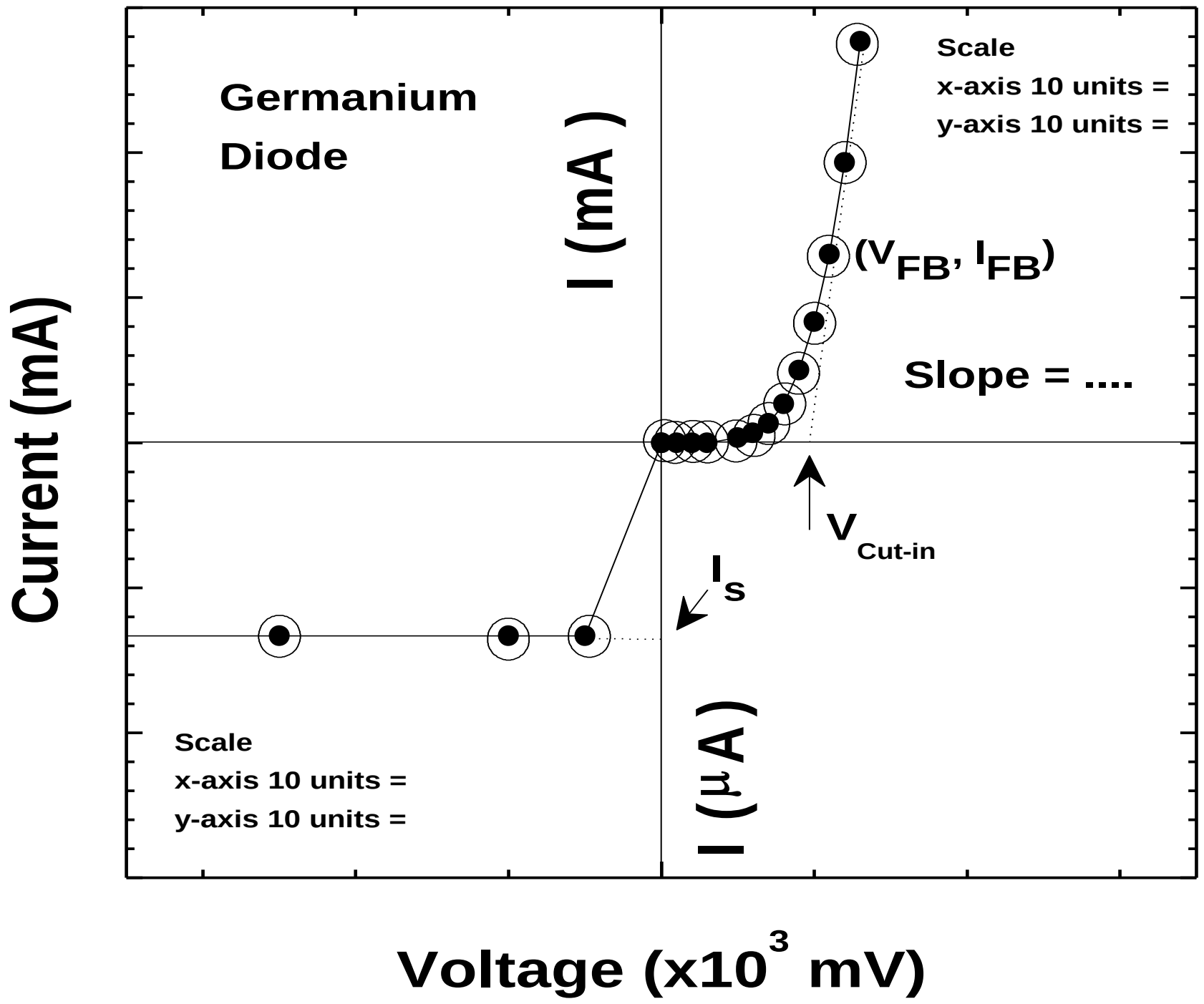

**b. Determination of the Energy gap of PN Junction Diode**

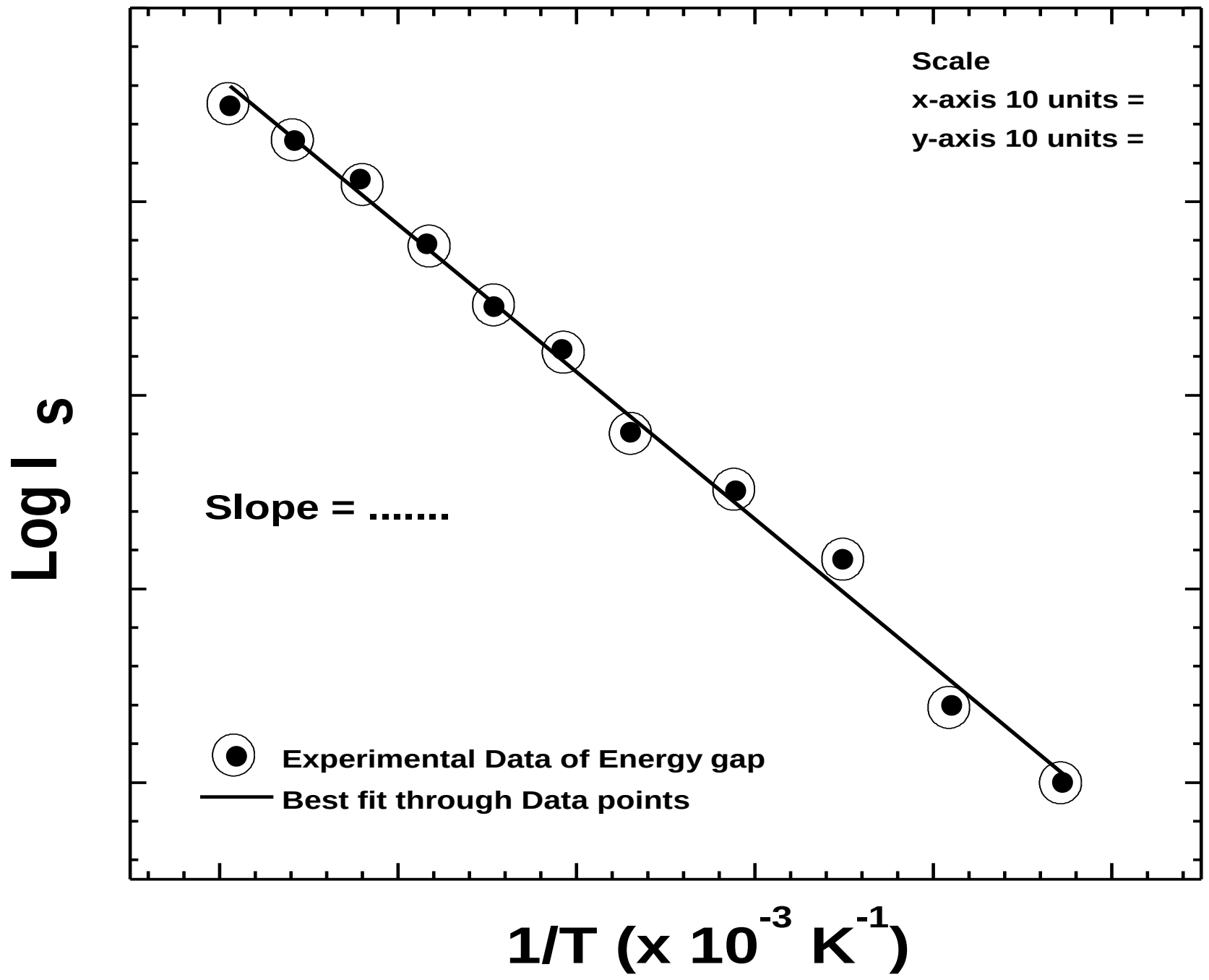

**6. Input and Output V-I Characteristics of the NPN transistor in Common Base Configuration**

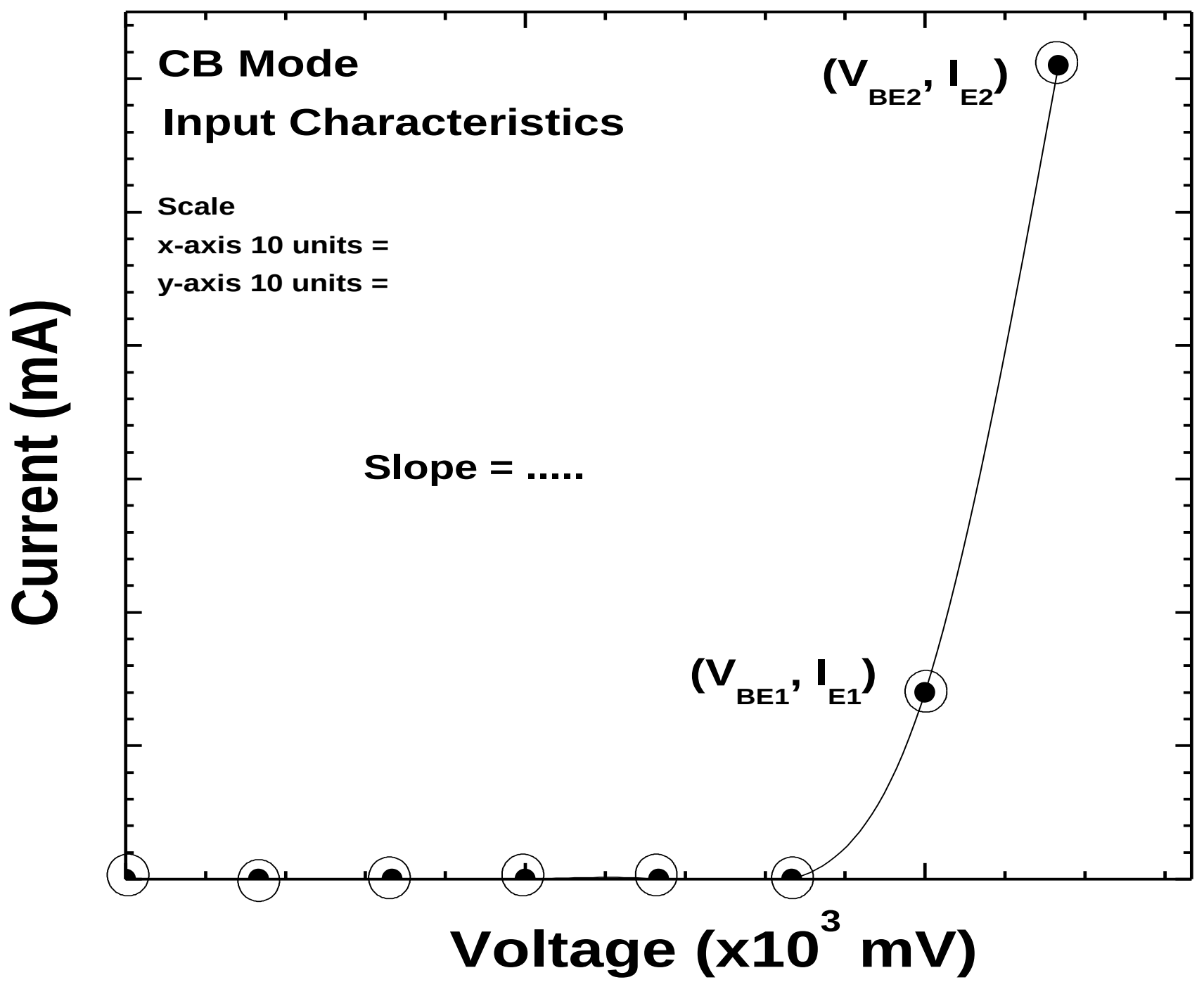

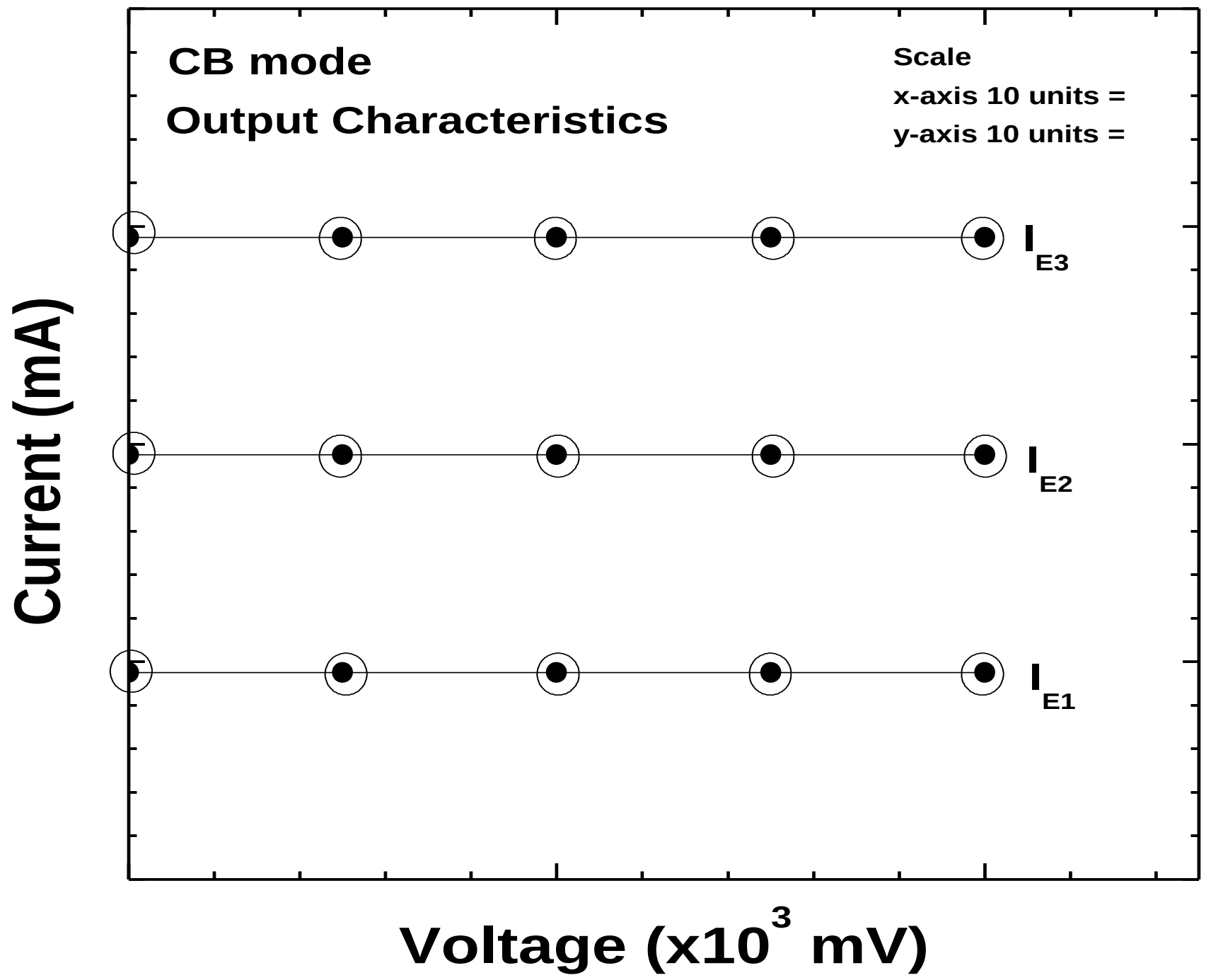

# 7a. Travelling Microscope

A travelling microscope is an optical instrument which is used to measure small distances and thus enables us to study various patterns formed due to interference of light. It mainly consists of a movable compound microscope attached with a scale that facilitates readings at different positions of the microscope.

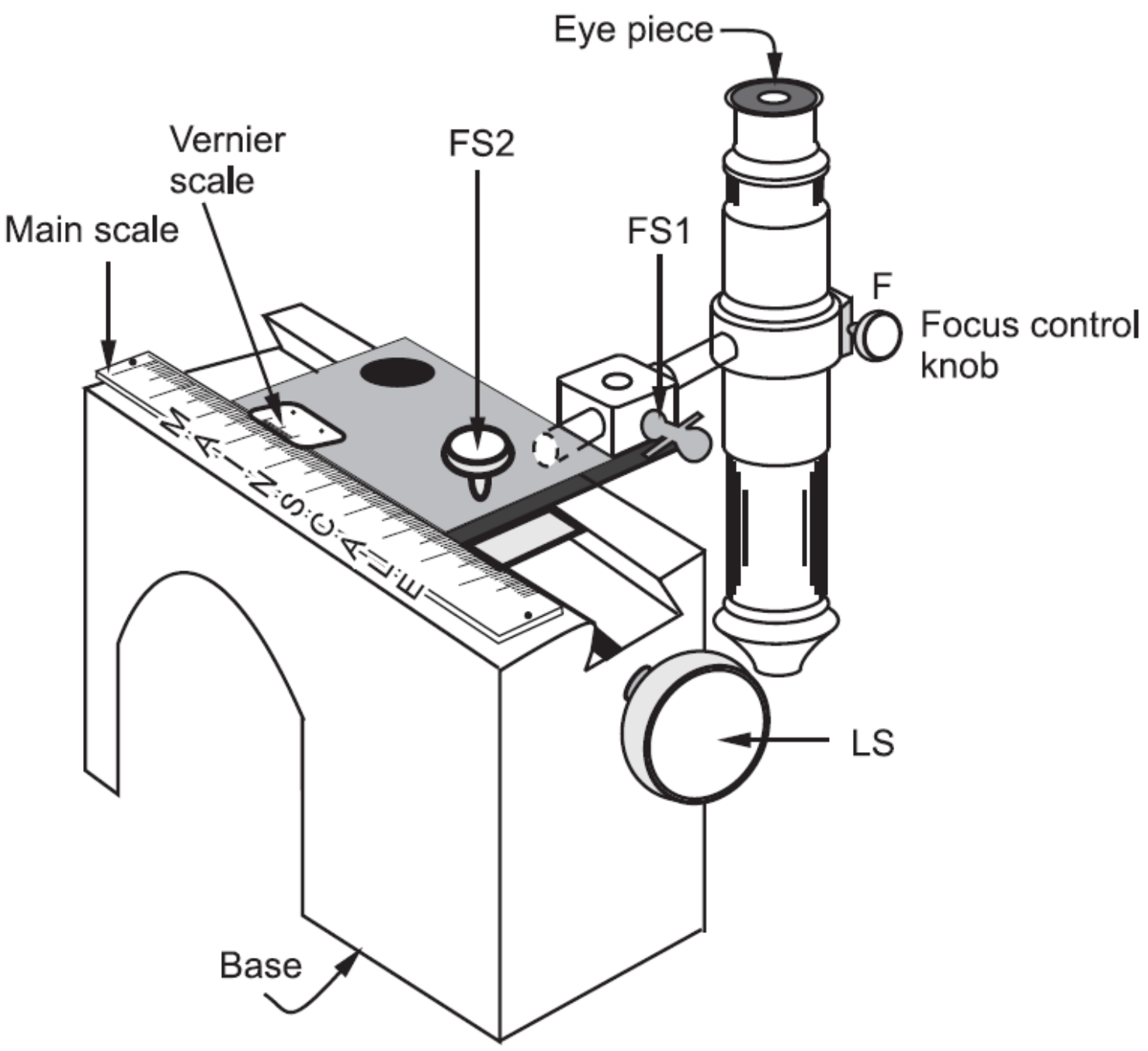

| Scale reading | Various adjustments |
|---|---|
| | |

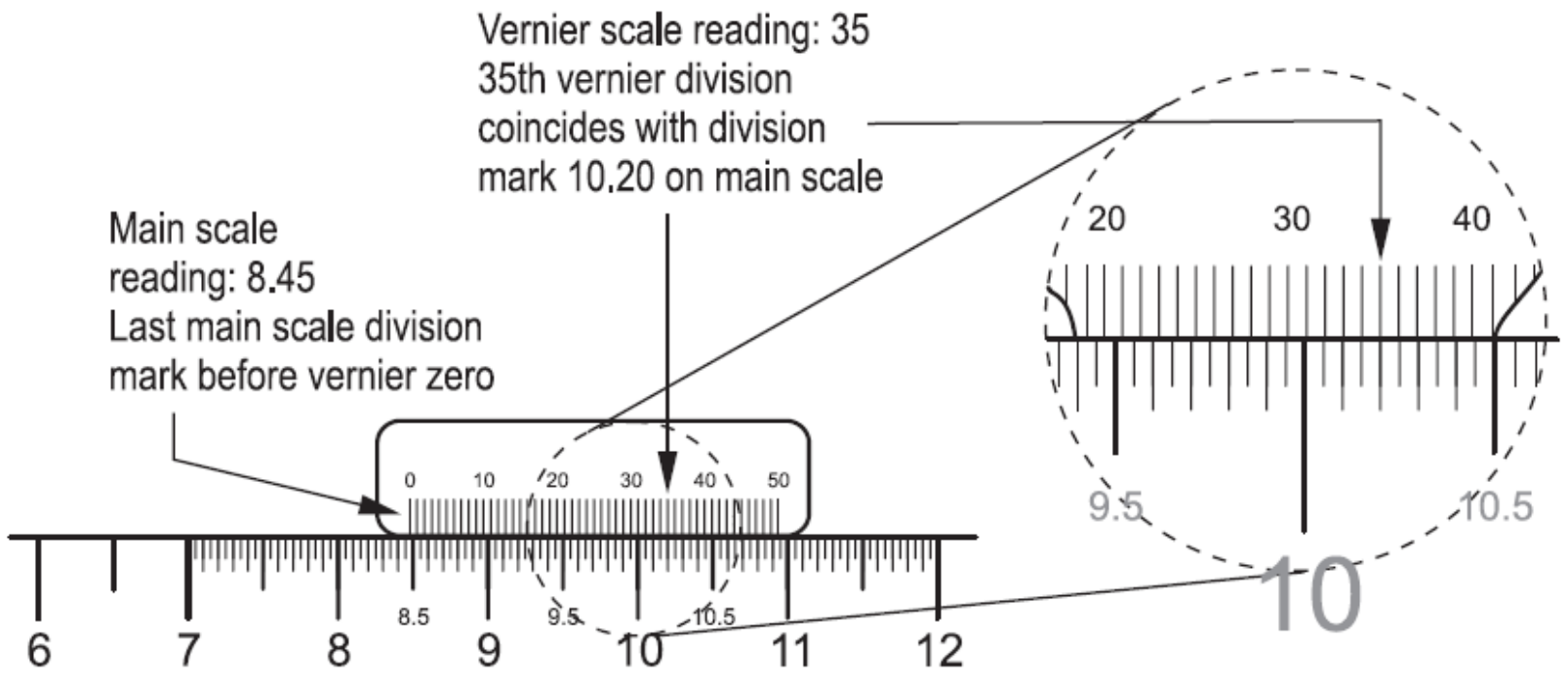

The method of taking readings and determination of least count is the same as explained in Chart 1.2. Thus for the case shown in the above figure,

Least count = Pitch/No. of divisions on the vernier scale
= 0.05 cm/50 = 0.001 cm

and TR = 8.45 cm + 35 × 0.001 cm = 8.485 cm

**Note:** The screw LS should always be moved in the same direction while taking readings. This is because reversing the direction of the screw introduces a *back lash error* which arises due to mechanical play in the threads of the screw.

1. The rack and pinion arrangement attached to the focus control knob F enables the adjustment of the distance between the object and the objective lens of the microscope, that is, it helps to bring the object into the focal plane of this lens.

2. FS1 and FS2 are the fixing screws. The screw FS2 is normally tightly closed so that the rotation of the long screw LS produces sliding of the vernier scale over the main scale. If larger distances are to be measured FS2 can be loosened and the scale can be mechanically slid by hand.

## 7b. Newton's Rings set-up

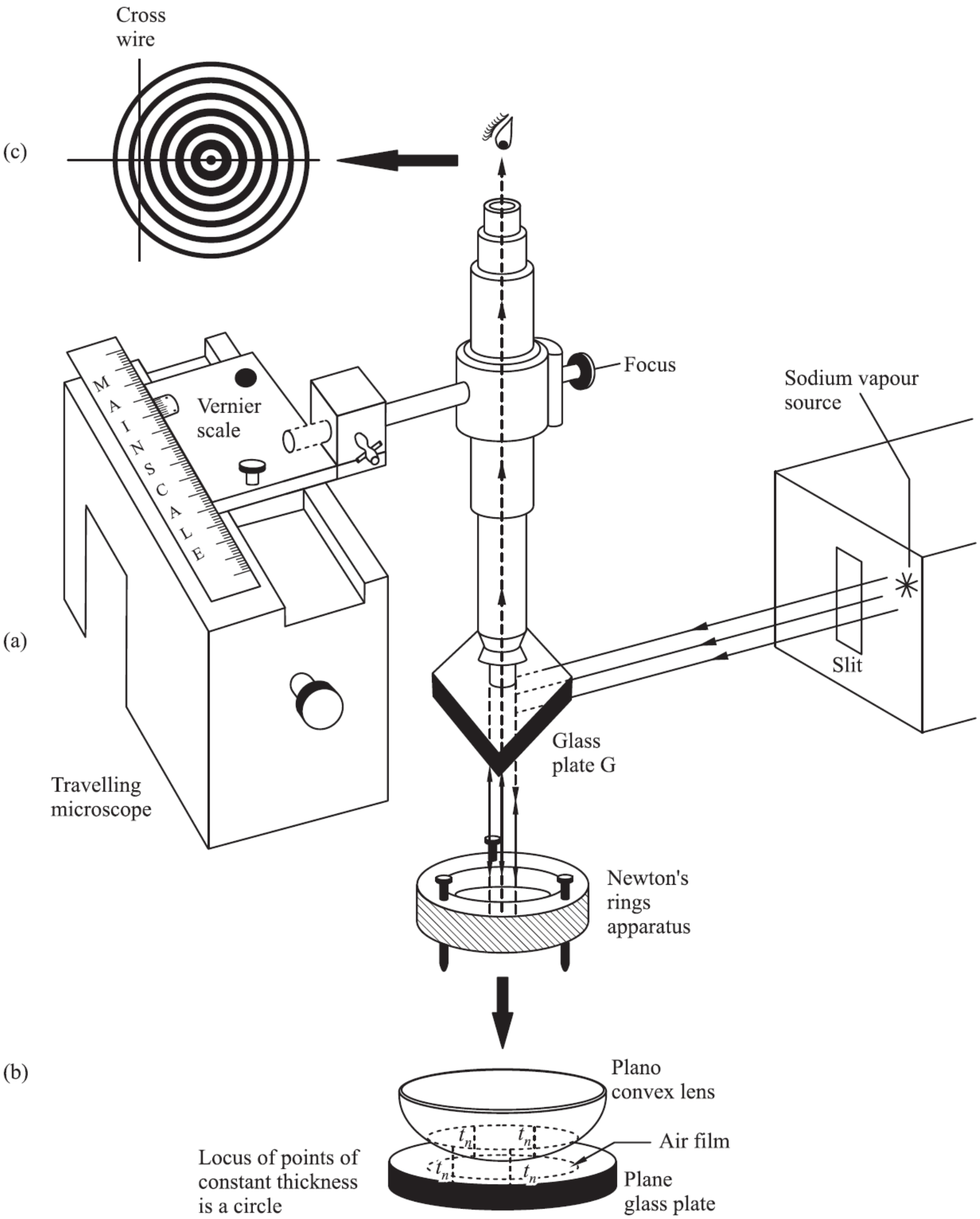

**Figure 3.17** Newton's rings: (a) Experimental arrangement, (b) Magnified view of air film and (c) Fringe pattern

**7c. Newton's Rings Graph**

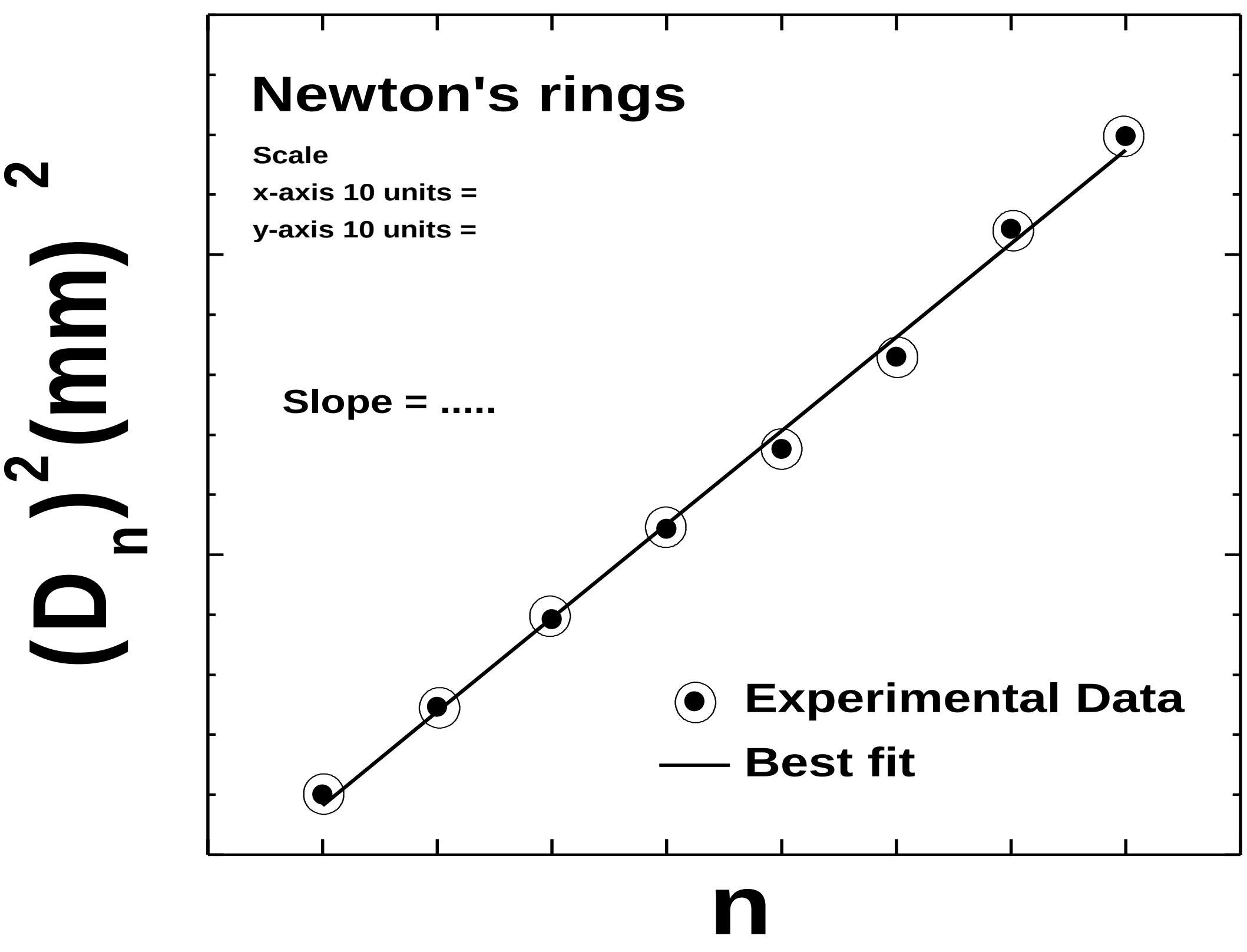

# 8a. Spectrometer

A spectrometer is an optical instrument which is used to measure angles and thus enables us to study phenomena involving the deviation / dispersion of light. A spectrometer mainly consists of a collimator for producing a parallel beam of light, a prism table for placing a prism or a grating and a telescope for observing the spectrum. The prism table and the telescope can be rotated and their angles of rotation can be measured using the circular scale attached to them. The prism table also has a facility for rotating without changing the reading of the scale. Both collimator and telescope have an objective lens, an eye piece and a focussing arrangement.

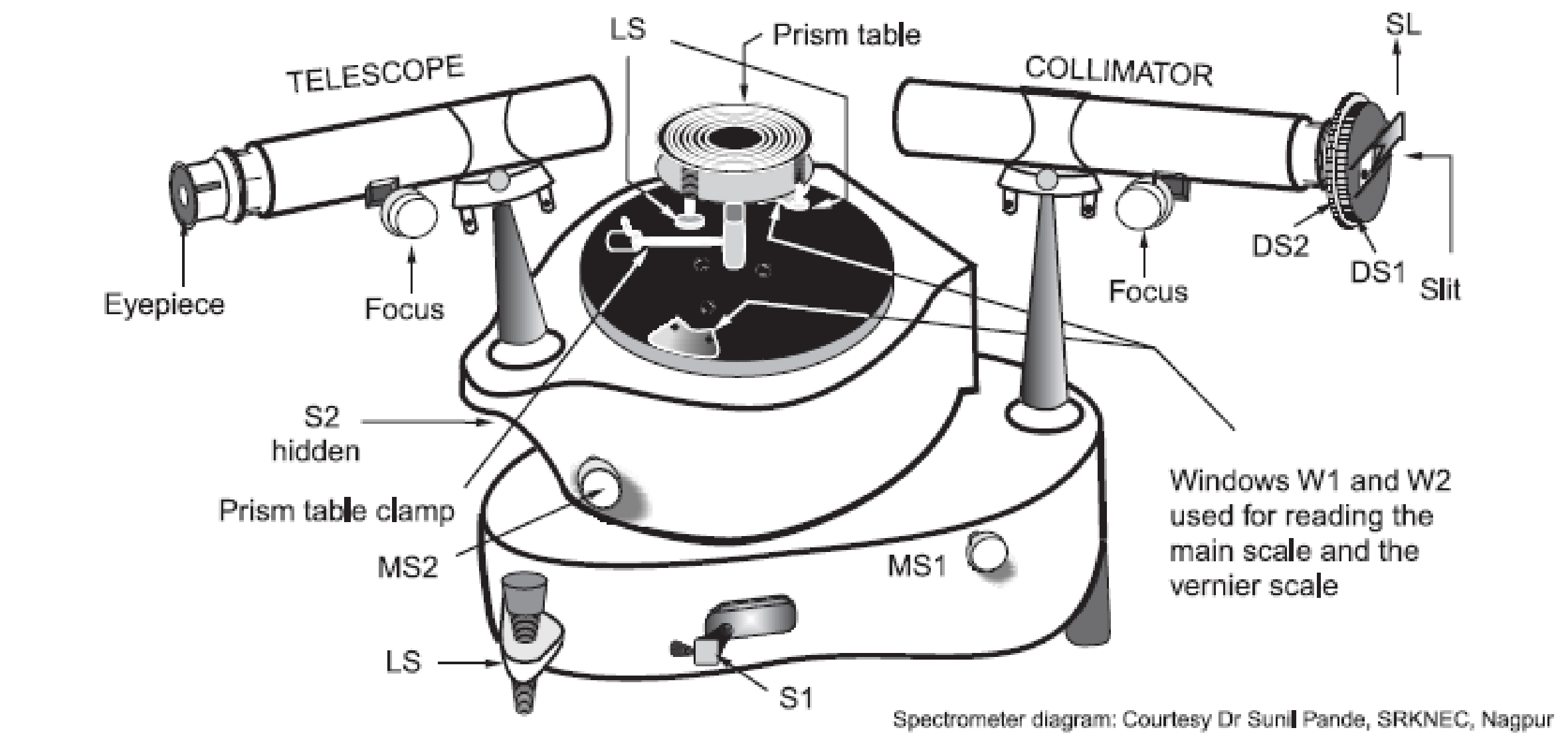

Spectrometer diagram: Courtesy Dr Sunil Pande, SRKNEC, Nagpur

**Window reading**

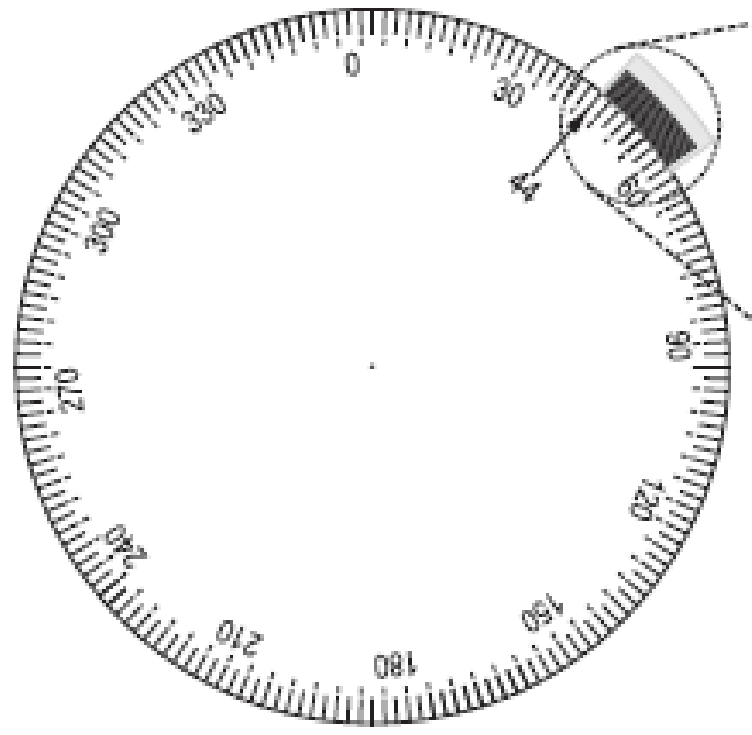
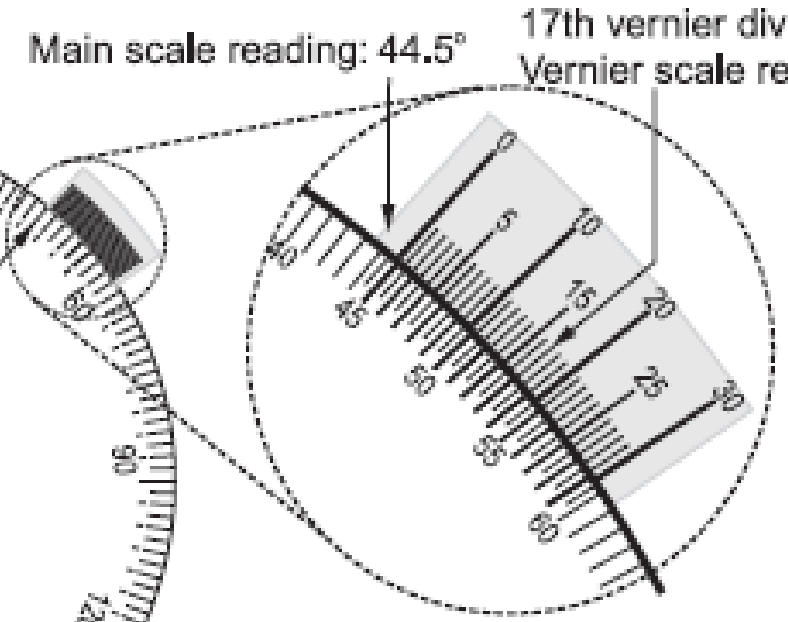

The method of taking readings and determination of least count is the same as explained in Chart 1.2. Thus for the case shown in the above figure,

Least count = pitch/No. of divisions on the vernier scale = (½)°/30 = (1/60)°
and TR = 44.5° + 17 × (1/60)° = 44.78°.

**NOTE**

1. Readings of $W_1$ and $W_2$ always differ by 180° ideally.
2. Maximum MSR is 360° which coincides with 0.
   This should be taken into consideration when scale passes through 360°. For example, if the telescope is rotated from an initial reading of 350° to the final reading of 10°, passing through the 360° mark, then the angle of rotation = (360 – 350) + (10 – 0) = 20°.

**Various adjustments**

1. The base of the spectrometer and the prism table are properly levelled using a spirit level and the screws LS provided for this purpose.
2. The orientation and the height of the slit are adjusted by the drum screw DS2 and the slide SL, respectively, to get a vertical slit of sufficient height. The slit width is adjusted using the drum screw DS1 to get a thin slit. However the intensity of light coming out of the slit is ensured to be sufficient for observation by positioning the collimator properly in front of the light source.
3. The collimation of a light beam, that is, rendering of an incident beam of light into a parallel beam is achieved by means of a rack and pinion arrangement provided with the collimator. This arrangement adjusts the distance between the slit and the collimating lens to bring the slit into the focal plane of this lens.
4. The collimator and the telescope are properly levelled so that the image of the slit is in the centre of the view.
5. If either the telescope or the windows are to be rotated by small angles, this can be done by locking their positions using S1 and S2 and using micromovement screws MS1 and MS2.

# 8b. Diffraction grating set-up

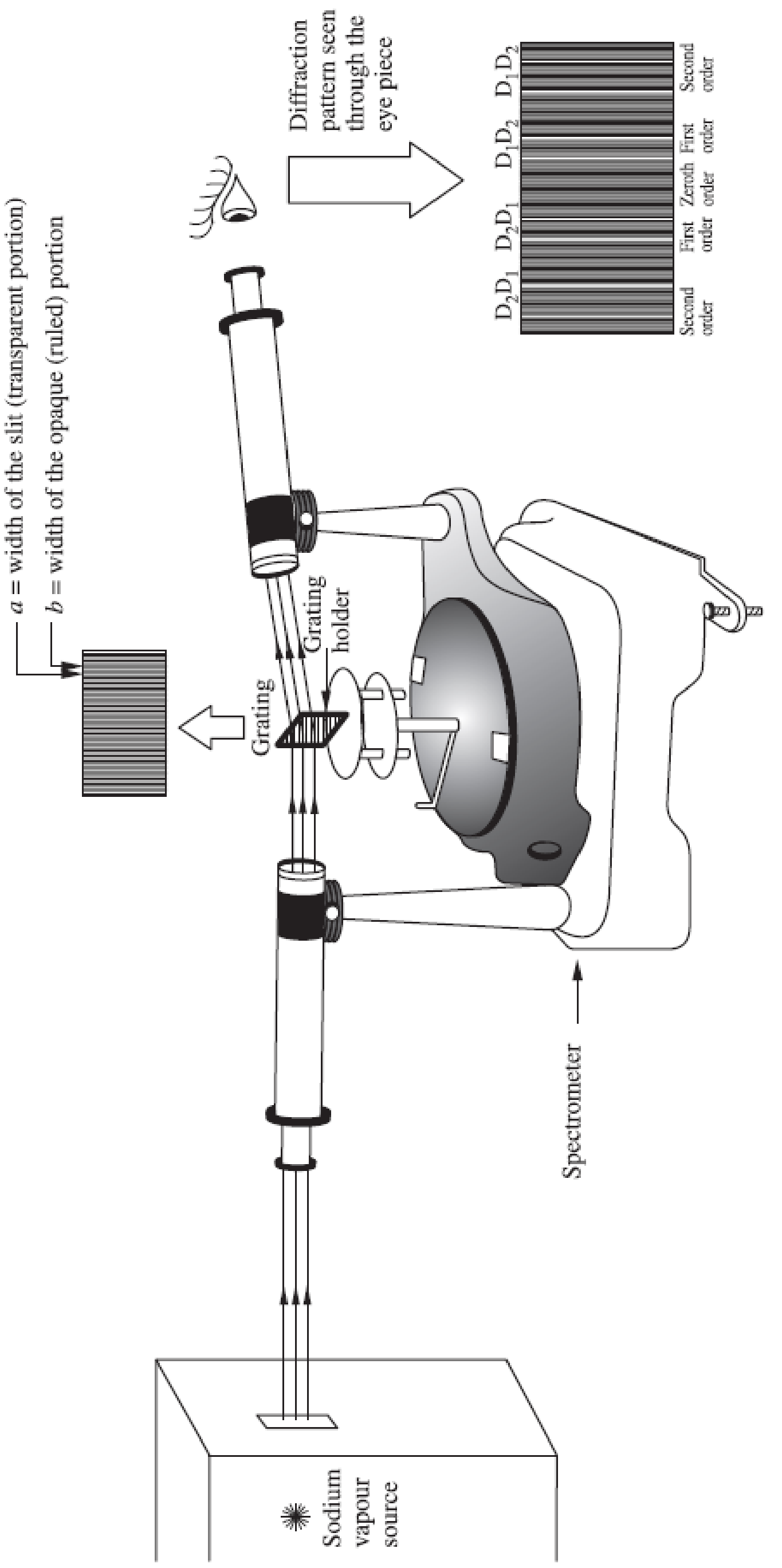

**Figure 4.7** Experimental arrangement to study diffraction of light

**Suppl. Method of Linear least square fitting**

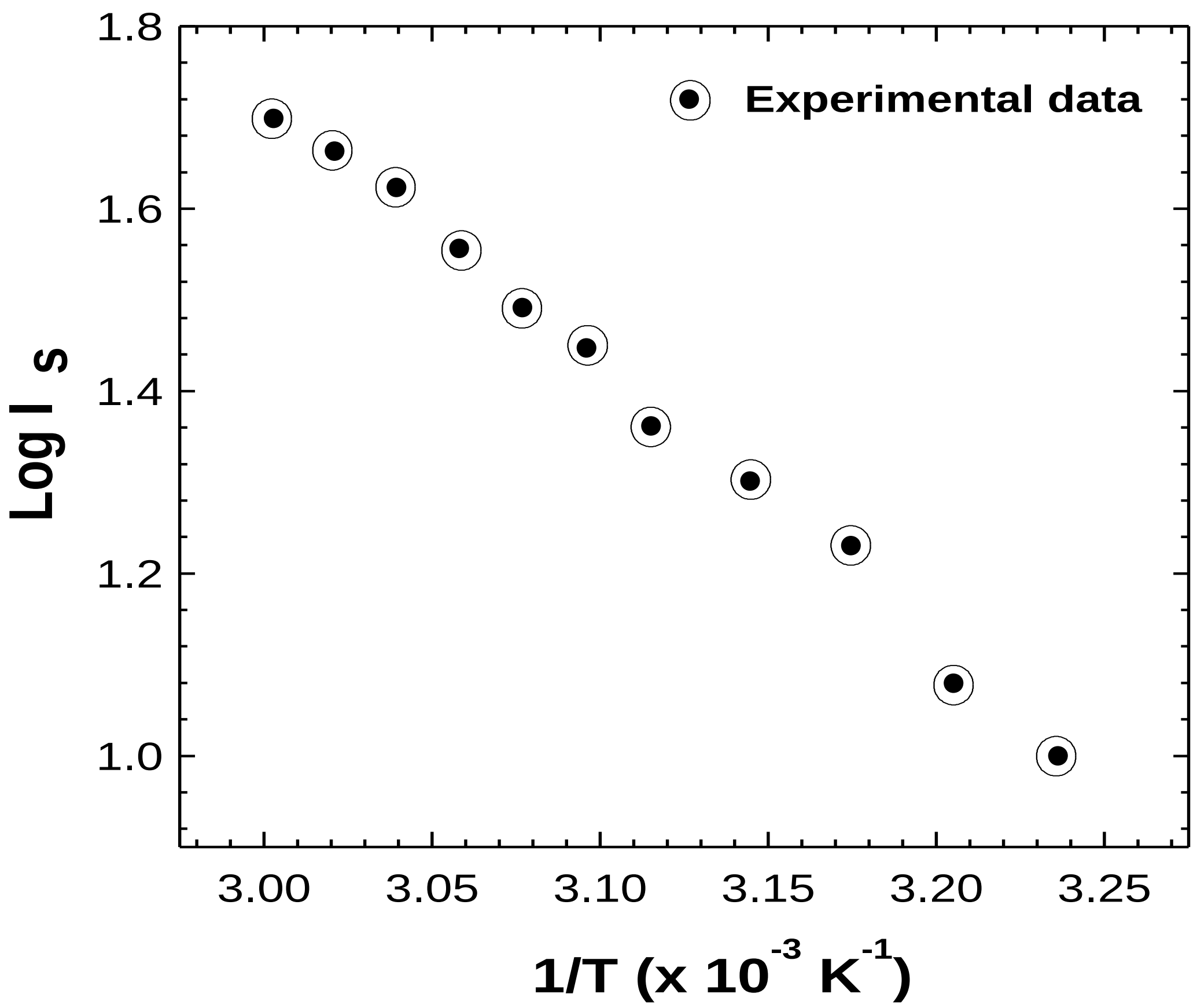

**Questions for Examination**

**Experiment 1. V-I Characteristics of the Photocell**

**1.** Plot I vs $1/d^2$ for V = 4 Volts for the given photocell.

Choose the distance (d) as the difference between the source window and the wooden window of the photocell. Take 5 data points.

Plot the best fit line for 5 points. Given the best parameters,

$m = [N \sum_i x_i y_i - \sum_i x_i \sum_i y_i ] / [N \sum_i x_i^2 - (\sum_i x_i)^2]$

$c = [\sum_i x_i^2 \sum_i y_i - \sum_i x_i \sum_i x_i y_i ] / [N \sum_i x_i^2 - (\sum_i x_i)^2]$

$y_{ls} = mx_i + c$

$E = \sum_i (y_i - y_{ls})^2$

Where, m = slope of the line of best fit,

c = intercept of the line of best fit,

$y_i$= the observed value of y at $x = x_i$,

$y_{ls}$= the corresponding value on the line of the best fit and

N = the total number of observations.

**2.** Calculate the work function and the error in the work function, i.e., $W_0 \pm \delta W_0$

To find the accuracy of the digital multimeter:

Complete accuracy specification: ±(% reading + number of LSD),

where,

Reading = true value of the signal measured by the Digital multimeter,

LSD = Least significant digit

Given, accuracy specification = ±(1.5% + 3),

Range of the Digital multimeter used, X.XXX

Accuracy of full range of 2 V = 1.999 ± [(1.999)(1.5)/100 + 0.003] volts

**Viva Questions**

1. What is a photocell? Why the cathode is cylindrical? What is the meaning of the workfunction of the cathode material?

2. Why the BG is used in this experiment? What is the least count of the BG?

3. How do you control the intensity of light incident on the photocell? What is the direction of current if the cathode is reverse biased?

4. If the same sodium source is used for interference experiment on one side and the photoelectric effect on other side, which experiment will work? Why?

5. On what factors the LSD given above depends?

**Experiment 2. Hall Effect**

**1.** Calculate the Hall coefficient $R_H$ from $V_H$ vs B for $I_s$ = 4 mA. Plot the best fit line for 6 points. Given the best parameters,

$m = [N \sum_i x_i y_i - \sum_i x_i \sum_i y_i ] / [N \sum_i x_i^2 - (\sum_i x_i)^2]$    $c = [\sum_i x_i^2 \sum_i y_i - \sum_i x_i \sum_i x_i y_i ] / [N \sum_i x_i^2 - (\sum_i x_i)^2]$

$y_{ls} = mx_i + c$  and  $E = \sum_i (y_i - y_{ls})^2$

Where, m = slope of the line of best fit,  c = intercept of the line of best fit,

$y_i$= the observed value of y at $x = x_i$,

$y_{ls}$= the corresponding value on the line of the best fit and

N = the total number of observations.

**2.** Calculate the Hall coefficient $R_H$ from $V_H$ vs $I_s$ for the source current through electromagnet, i.e., I = 1 A. Take 6 data points. Calculate $V_H \pm \delta V_H$ for all six points.

To find the accuracy of the digital multimeter:

Complete accuracy specification: ±(% reading + number of LSD),

where,

Reading = true value of the signal measured by the Digital multimeter,

LSD = Least significant digit

Given, accuracy specification = ±(1.5% + 3),

Range of the Digital multimeter used,  X.XXX

Accuracy of full range of 2 V = 1.999 ± [(1.999)(1.5)/100 + 0.003] volts

**Viva Questions**

1. What is Hall effect? What are the applications of Hall effect?

2. How is the magnetic field produced in the experiment? Does the strength of the magnetic field depend upon the distance between the polepieces?

3. What will happen to the voltage if the current through the crystal is reversed in its direction? What will happen if the direction of the applied magnetic field is reversed? What will happen if both the current and the magnetic field are reversed?

4. Can Hall effect be observed in conductors? Which of the materials (metals or semiconductors) have larger Hall coefficients? Why does Hall effect assume importance in semiconductors?

5. Will the Hall voltage vary with temperature in the metal and semiconductors? Justify your answer?

## Experiment 3. e/m by Thomson's method

Calculate e/m using Thomson's method

**Viva Questions**

1. Is a CRT solid state device? What are the principal parts of a CRT?

2. What is the function of an electron gun? What are the different parts of an electron gun?

3. At what potential is the grid kept with respect to the cathode? How does the grid control the intensity of the light spot on the screen?

4. How are the mutually repelling electrons brought to focus? How an electron beam is made visible in a CRT? How can the position of an electron beam be controlled?

5. What are deflection plates? How many sets of deflection plates are there in a CRT?

## Experiment 4. Cathode ray oscilloscope

Find the $V_{rms}$ and the average voltage for the sine wave, square wave and triangular wave.

Generate the required type of wave of 1 kHz from the function generator by pressing the Square wave button. Set the signal voltage using the level knob so that on the CRO screen the $V_{p-p}$ is 2 Volts.

**Viva Questions**

1. What is an electrical signal? Where is it applied in a CRO? What is a time-base and what is its importance?

2. How the profile of the signal is made visible on the screen? In the absence of time-base what would be the shape of trace on the screen?

3. Can a sinusoidal voltage be used as time base in place of a sawtooth voltage? If no, why? Why the sweep voltage is called time base and sawtooth voltage?

4. What is meant by blanking? Why is it necessary? What is meant by synchronization? How is synchronization achieved?

5. What is the function of volts/div control? What is meant by peak voltage? RMS voltage? Average voltage? Which voltage is measured by ordinary meters?

**Experiment 5a. PN Junction diode and Energy gap**

Find the static resistance $R_s$, the dynamic resistance $R_d$, $\Delta R_s$ and $\Delta R_d$ for the given pn-junction diode in the forward bias.

Use, (a) For static resistance of the PN Junction diode,

Given, $R_s = V_{FB}/I_{FB} = f(V, I)$

$$\Delta R_s = |\partial f / \partial V| \Delta V + |\partial f / \partial I| \Delta I$$

$$\Delta R_s = (1/I)\Delta V + (1/I^2)\, V\, \Delta I$$

(b) For dynamic resistance of the PN Junction diode,

$$R_d = (V_2 - V_1)/(I_2 - I_1)$$

$$\Delta R_d = 2\, \{[1/(I_2 - I_1)]\, \Delta V + [(V_2 - V_1)/(I_2 - I_1)^2]\, \Delta I\}$$

where, $\Delta V$ = (1/2) L.C. of the voltmeter and $\Delta I$ = (1/2) L.C. of the ammeter

**Viva Questions**

1. What is meant by biasing a device? What is meant by forward biasing the diode?

2. What is a pn junction? How it is formed?

3. What is the minimum value of forward bias voltage across a Germanium diode before it will conduct appreciably? On what factors the cut-in voltage depends?

4. What is the physical reason for the cut-in voltage?

5. What are the applications of a pn-junction diode?

**Experiment 5b. Energy gap**

Plot the following relation for the diode in the reverse bias at 4 V,

$$I_s = I_0 \exp(-E_g/kT).$$

Use, $E_g = 0.7$ eV and

The intercept, $c = [\sum_i x_i^2 \sum_i y_i - \sum_i x_i \sum_i x_i y_i] / [N \sum_i x_i^2 - (\sum_i x_i)^2]$

$y_i$= the observed value of y at $x = x_i$,

N = the total number of observations.

Take atleast 8 data points of temperature. (Possibly between 80 degrees to 55 degrees in the interval of 2 degrees).

**Viva Questions**

1. What is meant by band gap? Why is it called forbidden energy gap?

2. What happens in the valence band when an electron leaves it? Where does a free electron move? Where does a hole move?

3. What is an extrinsic semiconductor? How is it prepared? Does the band gap get affected by the addition of impurities?

4. What is the procedure adopted for determining the band gap in this experiment?

5. Why is the diode reverse biased in the experiment? Can the diode be forward biased to perform the measurement of the band gap?

**Experiment 6. Transistor**

**1.** Calculate the input resistance, $R_i$ and $\Delta R_i$ for the given NPN transistor in the common base configuration for $V_{CB}$ = 4 V.

Use $R_i = (V_{BE2} - V_{BE1})/(I_{E2} - I_{E1})$

$$\Delta R_i = 2 \{[1/(I_{E2} - I_{E1})] \Delta V + [(V_{BE2} - V_{BE1})/(I_{E2} - I_{E1})^2] \Delta I\}$$

where, $\Delta V$ = (1/2) L.C. of the voltmeter

$\Delta I$ = (1/2) L.C. of the ammeter

**2.** Plot the current transfer characteristics for the given NPN transistor in common base configuration. Calculate the current transfer gain $\alpha$ and $\Delta\alpha$ for $V_{CB}$ = 4 V.

Use $\Delta\alpha = 2[\Delta I_C / (I_{E2} - I_{E1}) + (I_{C2} - I_{C1}) \Delta I_E / (I_{E2} - I_{E1})^2]$

Where $\Delta I$ = (1/2) L.C. of the ammeter

**Viva Questions**

1. What is a transistor? How many regions are there in it? What are these regions called? What are the doping levels for these regions? Explain with reasons in short.

2. How many junctions form in a transistor? What are they called?

3. How are the junctions biased normally? Why a transistor has low input resistance and high output resistance? Why is the name transistor given to this device?

4. Draw the circuit for the common emitter configuration? What is the input voltage, input current, output voltage and output current in CE mode?

5. What is the base current due to? Why is the base current small in a transistor? How can a small base current control a large current in collector?

**Experiment 7. Newton's rings**

**1.** Calculate the wavelength $\lambda$ for R = 100 cm. Plot the best fit line for 5 points. Given the best parameters,

$m = [N \sum_i x_i y_i - \sum_i x_i \sum_i y_i ] / [N \sum_i x_i^2 - (\sum_i x_i)^2]$

$c = [\sum_i x_i^2 \sum_i y_i - \sum_i x_i \sum_i x_i y_i ] / [N \sum_i x_i^2 - (\sum_i x_i)^2]$

$y_{ls} = mx_i + c$

$E = \sum_i (y_i - y_{ls})^2$

Where, m = slope of the line of best fit,

c = intercept of the line of best fit,

$y_i$= the observed value of y at $x = x_i$,

$y_{ls}$= the corresponding value on the line of the best fit and

N = the total number of observations.

**2.** Find the area of circles formed by the first five dark rings. Plot the area as a function of ring number. Calculate the wavelength $\lambda$ for R = 100 cm. Plot the best fit line for 5 points. Given the best parameters,

$m = [N \sum_i x_i y_i - \sum_i x_i \sum_i y_i] / [N \sum_i x_i^2 - (\sum_i x_i)^2]$

$c = [\sum_i x_i^2 \sum_i y_i - \sum_i x_i \sum_i x_i y_i] / [N \sum_i x_i^2 - (\sum_i x_i)^2]$

$y_{ls} = m x_i + c$

$E = \sum_i (y_i - y_{ls})^2$

Where, m = slope of the line of best fit,

c = intercept of the line of best fit,

$y_i$= the observed value of y at $x = x_i$,

$y_{ls}$= the corresponding value on the line of the best fit and

N = the total number of observations.

**Viva Questions**

1. Why are the Newton's rings circular and concentric? Why is the central spot dark?

2. Why do the rings get closer as the order of the fringe increases? On what factors does the diameter of a ring depend?

3. How will the rings change if we introduce a little water between the lens and the plate?

4. What is the role of the inclined glass sheet kept in the part of light? At what angle should be arranged?

5. What are the engineering applications of Newton's rings?

**Experiment 8. Diffraction**

**1.** Determine the angular separation between the spectral lines D1 and D2 for the left hand side and the right hand side. Compare the experimental value with the theoretical value using the formula.

Use n = 2, i.e, the $2^{nd}$ order diffraction spectra on both sides.

Given $\lambda_1$ = 5890 Angstroms and $\lambda_2$ = 5896 Angstroms

**2.** Calculate the angular separation between the direct reading and the grating surface at 53.5° with respect to the collimator.

**Viva Questions**

1. What is meant by diffraction of light? Why is diffraction of light not noticeable in everyday life?

2. What is a diffraction grating? How is it prepared? What is grating spacing? What is the relationship between the grating spacing and the number of rulings per cm on the grating?

3. Explain with diagram how grating forms diffraction images when monochromatic light is incident on it?

4. What is meant by the order of the spectrum? Where is the zero-order fringe formed?

5. Do the spectra of different orders have the same intensity? Do you find any difference in the separation of lines in the various orders?